\documentclass{aastex61}

\usepackage{amsmath}            
\usepackage{multirow}
\usepackage{listings}
\usepackage{xcolor}                  
\usepackage{longtable}

\newcommand{\Msun}{\,\mbox{$M_{\odot}$}\,}

\newcommand{\eg}{e.g.\,\,}
\newcommand{\ie}{i.e.\,\,}

\newcommand{\Teff}{\mbox{$T_{\rm eff}$}\,}

\newcommand{\Myr}{\mbox{$Myr$}\,}

\newcommand{\masyr}{\,mas\,\mbox{$yr^{-1}$}\,}

\newcommand{\WaWb}{\mbox{$(W1-W2)$}\,}

\newcommand{\HKcut}{\,{H}\kern-.1em{K}\,}
\newcommand{\Redcut}{\,\mbox{$R$}\,}
\newcommand{\noHKcut}{}
\newcommand{\noRedcut}{}
\newcommand{\UpperSco}{Upper Scorpius\,}

\newcommand{\Figure}{Figure }
\newcommand{\Table}{Table }
\newcommand{\Section}{Section }
\newcommand{\Appendix}{Appendix }
\newcommand{\Equation}{Equation }

\definecolor{codegreen}{rgb}{0,0.6,0}
\definecolor{codegray}{rgb}{0.5,0.5,0.5}
\definecolor{codepurple}{rgb}{0.58,0,0.82}
\definecolor{backcolor}{rgb}{0.95,0.95,0.92}
\lstdefinestyle{mystyle}{
    basicstyle=\scriptsize,
    backgroundcolor=\color{backcolor},   
    commentstyle=\color{codegreen},
    keywordstyle=\color{orange},
    stringstyle=\color{codepurple},
    breakatwhitespace=false,         
    breaklines=true,                 
    captionpos=b,                    
    keepspaces=true,                                 
    showspaces=false,                
    showstringspaces=false,
    showtabs=false,                  
    tabsize=1,
}
\received{September 11, 2017}
\revised{October 16, 2017}
\accepted{October 17, 2017}
\submitjournal{AJ}

\shorttitle{VLMOs in USco using Gaia DR1}
\shortauthors{Cook et al.}

\begin{document}

\title{Very Low-Mass Stars and Brown Dwarfs in Upper Scorpius using Gaia DR1: Mass Function, Disks and Kinematics}

\correspondingauthor{Neil J. Cook}
\email{neil.james.cook@gmail.com}

\author[0000-0003-4166-4121]{Neil J. Cook}
\affil{Institut de Recherche sur les Exoplan\`{e}tes, Universit\'{e} de Montr\'{e}al, Montr\'{e}al, QC, Canada, H3T 1J4}
\affil{Faculty of Science, York University, 4700 Keele Street, Toronto, Canada, ON M3J 1P3}

\author[0000-0001-8993-5053]{Aleks Scholz}
\affil{SUPA, School of Physics \& Astronomy, University of St. Andrews, North Haugh, St. Andrews, KY16 9SS, UK}

\author[0000-0001-5349-6853]{Ray Jayawardhana}
\affil{Faculty of Science, York University, 4700 Keele Street, Toronto, Canada, ON M3J 1P3}



\begin{abstract}
Our understanding of the brown dwarf population in star forming regions is dependent on knowing distances and proper motions, and therefore will be improved through the {\itshape Gaia} space mission. In this paper, we select new samples of very low mass objects (VLMOs) in Upper Scorpius using UKIDSS colors and optimised proper motions calculated using Gaia DR1. The scatter in proper motions from VLMOs in Upper Scorpius is now (for the first time) dominated by the kinematic spread of the region itself, not by the positional uncertainties. With age and mass estimates updated using Gaia parallaxes for early type stars in the same region, we determine masses for all VLMOs. Our final most complete sample includes 453 VLMOs of which $\sim$125 are expected to be brown dwarfs. The cleanest sample is comprised of 131 VLMOs, with $\sim$105 brown dwarfs. We also compile a joint sample from the literature which includes 415 VLMOs, out of which 152 are likely brown dwarfs. The disc fraction among low-mass brown dwarfs ($M<0.05 \Msun$) is substantially higher than in more massive objects, indicating that discs around low-mass brown dwarfs survive longer than in low-mass stars overall. The mass function for $0.01<M<0.1 \Msun$ is consistent with the Kroupa IMF. We investigate the possibility that some `proper motion outliers' have undergone a dynamical ejection early in their evolution. Our analysis shows that the color-magnitude cuts used when selecting samples introduce strong bias into the population statistics due to varying level of contamination and completeness.
\end{abstract}




\section{Introduction}
\label{section:intro}

Most newly formed stars have masses significantly lower than the Sun. The characteristic mass of star formation, the peak of the Initial Mass Function (IMF), is around 0.2$\,\Msun$, almost independent of environment \citep{Bonnell2007}. The mass distribution of objects formed in young clusters extends far below the sub-stellar limit at 0.08$\Msun$ and into the planetary (fusion-less) mass domain at $<0.015\Msun$ \citep{Luhman2012,Scholz2012}. In this very low mass (VLM) domain, a variety of formation channels might play a role, including turbulent fragmentation of clouds, dynamical ejections from multiple systems, or disc fragmentation \citep[see][]{Whitwort2007}.

Very low mass object (VLMOs) are also viable host stars for exoplanet systems, as evidenced by discoveries of Earth-sized or -massed planets around mid M dwarfs \citep{BertaThompson2015,AngladaEscude2016,Dittmann2017,Gillon2017}. The ubiquity of these systems poses an interesting challenge for core accretion theories, as discs around young objects in this mass domain usually do not seem to have sufficient material to form these type of systems \citep{Scholz2006,Testi2015,Pascucc2016}, implying very rapid formation. VLMOs are also possible hosts of ultracool dwarfs with L, T, or Y spectral types \citep{Bardalez2013,Cook2016,Cook2017}. 

Identifying and characterizing the VLM population in star forming regions provides the observational constraints on star formation scenarios as well as the samples for in-situ studies of planet formation. Traditionally, the selection of VLM cluster members is based on cuts in color-magnitude and proper motion space, followed by spectroscopy to confirm \citep{Luhman2003,Wilking2004,Scholz2012}. So far, proper motion cuts were limited to a few nearby regions with space motions significantly offset from the Galactic background. The output from the astrometry mission {\itshape Gaia} is about to change that \citep{Gaia2016a,Gaia2016b}. It is anticipated that the final Gaia data releases will provide the first large, uniform sample of parallaxes for young brown dwarfs in addition to sub-milliarcsecond precision in proper motions.

Gaia published its first data release in 2017 \citep[][henceforth Gaia DR1]{Gaia2016a,Gaia2016b}. While Gaia DR1 does not yet provide Gaia-internal parallaxes and has not yet reached optimum astrometric precision, it can already be used for improving current selection methods for young VLMOs and to refine the resulting samples, as we demonstrate in this paper for the nearest OB association Upper Scorpius. Using the combined parallaxes from the {\itshape Tycho-Gaia Astrometric Solution} (TGAS) for bright young stars, the estimates for distances, age and spatial depth for nearby star forming regions can be solidified. With the help of the Gaia DR1 astrometry, the scatter in proper motions from VLMOS in \UpperSco is now (for the first time) dominated by the kinematic spread of the region itself, not by the positional uncertainties. For this particular region later data releases are unlikely to significantly improve the member selection from proper motions. \UpperSco is a region mostly free from reddening, therefore follow-up spectroscopy is not as essential here as in other star forming regions.

In this paper, we estimate distance, age and spatial depth from higher-mass \UpperSco members (\Section\ref{section:upper_sco}). We establish new samples of VLM members of Upper Scorpius using photometry from the {\itshape United Kingdom Infrared Digital Sky Survey Galactic Clusters Survey} \citep[UKIDSS/GCS, ][see \Section\ref{section:bds_in_upper_sco}]{Lawrence2007}, and optimized astrometry obtained by combining Gaia DR1 with other catalogs (\Section\ref{section:pm_analysis}). Using these new samples we estimate masses (\Section\ref{section:properties:isofit}), test the the disc fraction as a function of mass (\Section\ref{section:properties:discs}) and the mass function (\Section\ref{section:properties:mass_function}). We also make a first attempt at tracking down kinematic outliers (\Section\ref{section:properties:outliers}), \ie objects with proper motions significantly different from their nearby siblings, which could be those who experienced an early dynamical ejection from a disc. The paper is intended to pave the way for future studies of other star forming regions based on the next Gaia data releases.

For the purposes of this paper, we use the term VLMO for all objects with masses below $\lesssim 0.2\Msun$, including very low mass stars, brown dwarfs and planetary mass objects.

\section{The Upper Scorpius Association with Gaia DR1}
\label{section:upper_sco} 

\begin{figure*}
\begin{center}
\includegraphics[width=\textwidth]{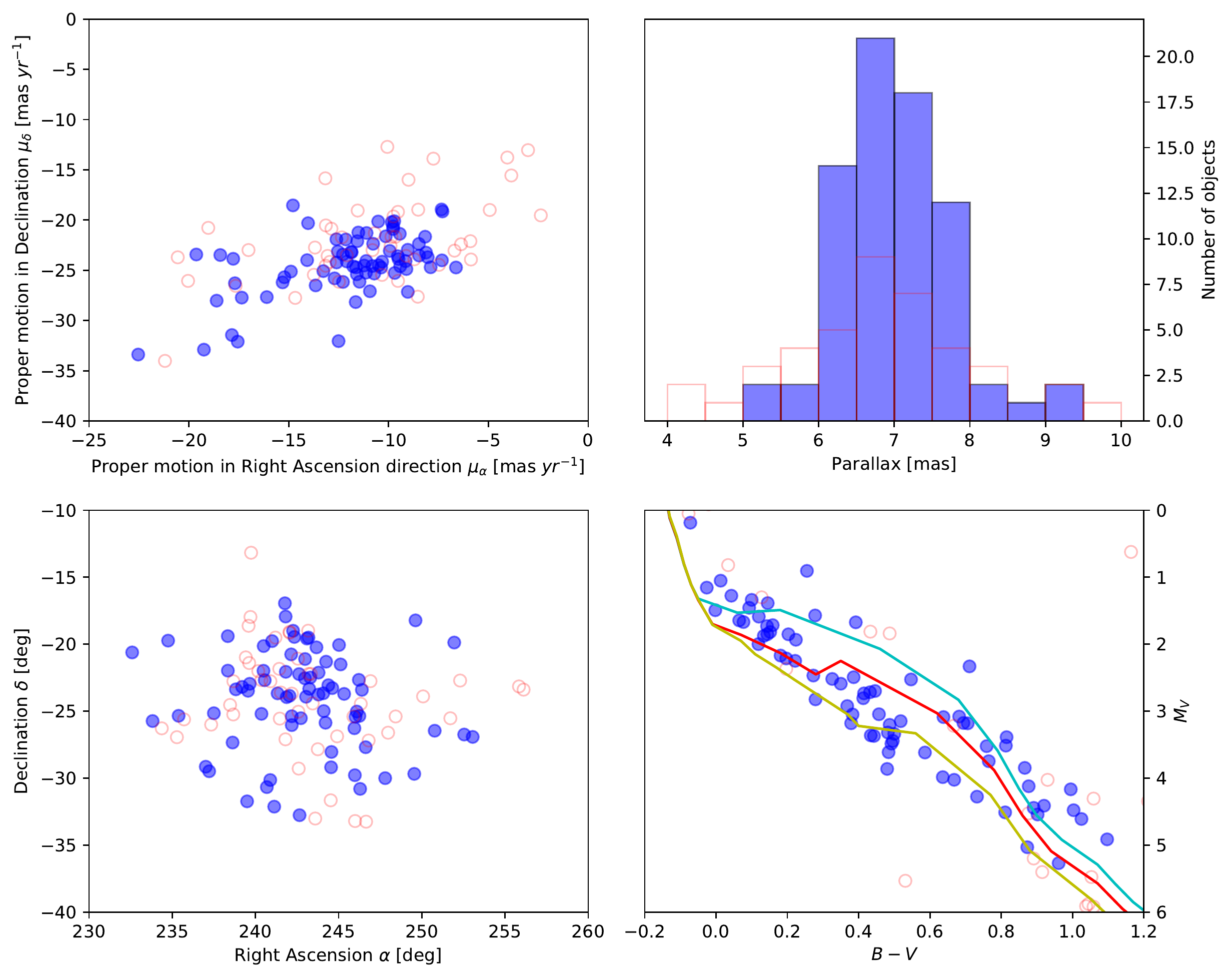}
\caption{The 74 higher-mass members of \UpperSco used to determine distance, age and spatial depth for \UpperSco (in blue). The upper left panel shows their distribution in proper motion, the upper right panel shows their distance spread, the bottom left panel shows the distribution in space, and the lower right panel shows the color-magnitude diagram with 7, 10, and 15\,\Myr isochrones$^{\ref{footnote:stellar-isochrones}}$ (cyan, red, yellow) from \protect\citet{Siess2000}. Over-plotted in red are the rejected candidates. \label{figure:section_2_pm_plot}}
\end{center}
\end{figure*}

When deriving stellar properties for members of star forming regions, the main source of uncertainty are distance and age. Gaia DR1 does not provide parallaxes for very low mass members of \UpperSco -- this is expected to be included in later data releases -- but the TGAS catalog does contain parallaxes for a substantial sample of early-type stellar members. TGAS\footnote{Documentation available from \url{https://gaia.esac.esa.int/documentation/GDR1/Data_processing/chap_cu3tyc/}}, is a combination of the Tycho-2 with the Gaia catalog, listing astrometric data for $\sim$2.5 million stars \citep{Michalik2014,Gaia2016a,Gaia2016b}. In this section we aim to use this dataset to re-determine distance and age for \UpperSco. 

We start with the sample of high-mass members of \UpperSco from \citet{deZeeuw1999}, selected as a moving group using Hipparcos astrometry. The mean Hipparcos distance in this sample is 145$\pm$1 pc. From the 120 members listed there, 85 have an entry in TGAS (with 1\arcsec matching radius). From the catalog of primarily K-type members by \citet{Pecaut2012}, we add 36 stars without Hipparcos, but with TGAS entry. To clean the sample, we remove objects with parallax error $>10\%$ (16 objects), without Tycho-2 photometry (4 objects) or magnitude error $>0.1$\,mag (24 objects), and with implausible distances $>200$\,pc (9 objects) leaving 74 objects (47 objects rejected in total, taking into account those rejected by multiple criteria).

Five of the rejected stars at distance $>200$\,pc are classified as K-M giants in the Michigan Spectral Survey \citep{Houk1988}, among them HIP83542, also rejected by \citet{Pecaut2012}. In \Figure\ref{figure:section_2_pm_plot}, the cleaned sample has clusters around $(-11.7,-24.2)$\,\masyr, with standard deviation of 3\masyr.

The cleaned sample of 74 stars have a median TGAS distance of 146.1\,pc, and a mean TGAS distance of 145$\pm$2\,pc. The median parallax error in this sample is 0.36\,mas -- this includes a systematic error of 0.14\,mas, see Table 2 in \citet{Arenou2017}. The parallax error translates to a median distance error of 7.8\,pc and a $\sqrt{N}$ scaled error of $\pm 0.9$\,pc for the distance of the association. Thus, this provisional distance estimate from TGAS is within the error bars of the Hipparcos result. The parallax distribution (see \Figure\ref{figure:section_2_pm_plot}, upper right panel) has a standard deviation of 15.2\,pc. Subtracting the errors in quadrature, this suggests a depth of the region along the line of sight of about $\pm 13$\,pc. For comparison, in the plane of the sky, the sample is contained within an area of 20\,deg diameter in both RA and DEC, corresponding to 50\,pc at the distance of \UpperSco. 

From the cleaned sample, we produce a color-magnitude diagram (see \Figure\ref{figure:section_2_pm_plot}, lower right panel). The absolute magnitudes have been calculated using the individual TGAS distances for each star, eliminating the error caused by the depth of the region. Uncertainties in $M_v$ and $B-V$ are comparable to the size of the symbols for most objects. Variability is expected to have a minor effect on the position of the stars in the diagram for a region of this age. Without widespread accretion or discs \citep{Luhman2012}, the only plausible source of variability is magnetic activity, which typically does not cause large amplitudes, even at this young age \citep{Grankin2008}. Therefore, the spread in the color-magnitude diagram likely corresponds to a real age spread in \UpperSco.

Over-plotted in \Figure\ref{figure:section_2_pm_plot} are the 7, 10, and 15\,\Myr isochrones\footnote{\url{http://www.astro.ulb.ac.be/~siess/pmwiki/pmwiki.php/WWWTools/Isochrones}\label{footnote:stellar-isochrones}} (cyan, red, yellow) from \citet{Siess2000}, using a metallicity of $Z=0.02$ (solar metallicity, \citealt{Siess2000}) and the \citet{Kenyon1995} conversion table. These three isochrones bracket most of the sample for $(B-V)<0.8$. For later-type objects the spread in the data points exceeds the model dispersion, partly due to increased photometric errors. Thus 7-15\,\Myr is a plausible minimum estimate for the age spread in  this region. This is consistent with recent age studies for \UpperSco, see \protect\citet{Pecaut2016,Fang2017}, but in conflict with earlier claims of 5$\pm$ 2\,\Myr by \citet{Preibisch2002}. Note that we checked isochrones with non-solar metallicity and the differences were negligible for our purposes.

\section{Very low mass objects in Upper Scorpius}
\label{section:bds_in_upper_sco}

Many previous works have studied the population of VLMOs in \UpperSco (\eg \citealt{Slesnick2008}; \citealt{Dawson2011}; \citealt{Lodieu2011}; \citealt{Luhman2012b}; \citealt{Dawson2013}; \citealt{Lodieu2013a}; \citealt{Lodieu2013b}; \citealt{Dawson2014}). Most use various color cuts to select potential VLMOs, and use proper motions (either calculated or obtained from large catalogs) to identify candidates moving in a similar way to known \UpperSco members. One of the main uncertainties of membership (when distance is unknown) is the uncertainties associated with the calculated proper motions. Thus more precise proper motions tend to lead to identifying a better sample of VLMOs in \UpperSco.

In this section we compile two samples, the first is a sample directly taken from the literature (and thus based on various colors cuts and slightly different selection criteria), henceforth the `L-sample', the second is a large uniform sample, based on the initial color selection from \citet{Lodieu2013a}, henceforth the `C-sample'.

\begin{figure*}
\begin{center}
\includegraphics[width=.95\textwidth]{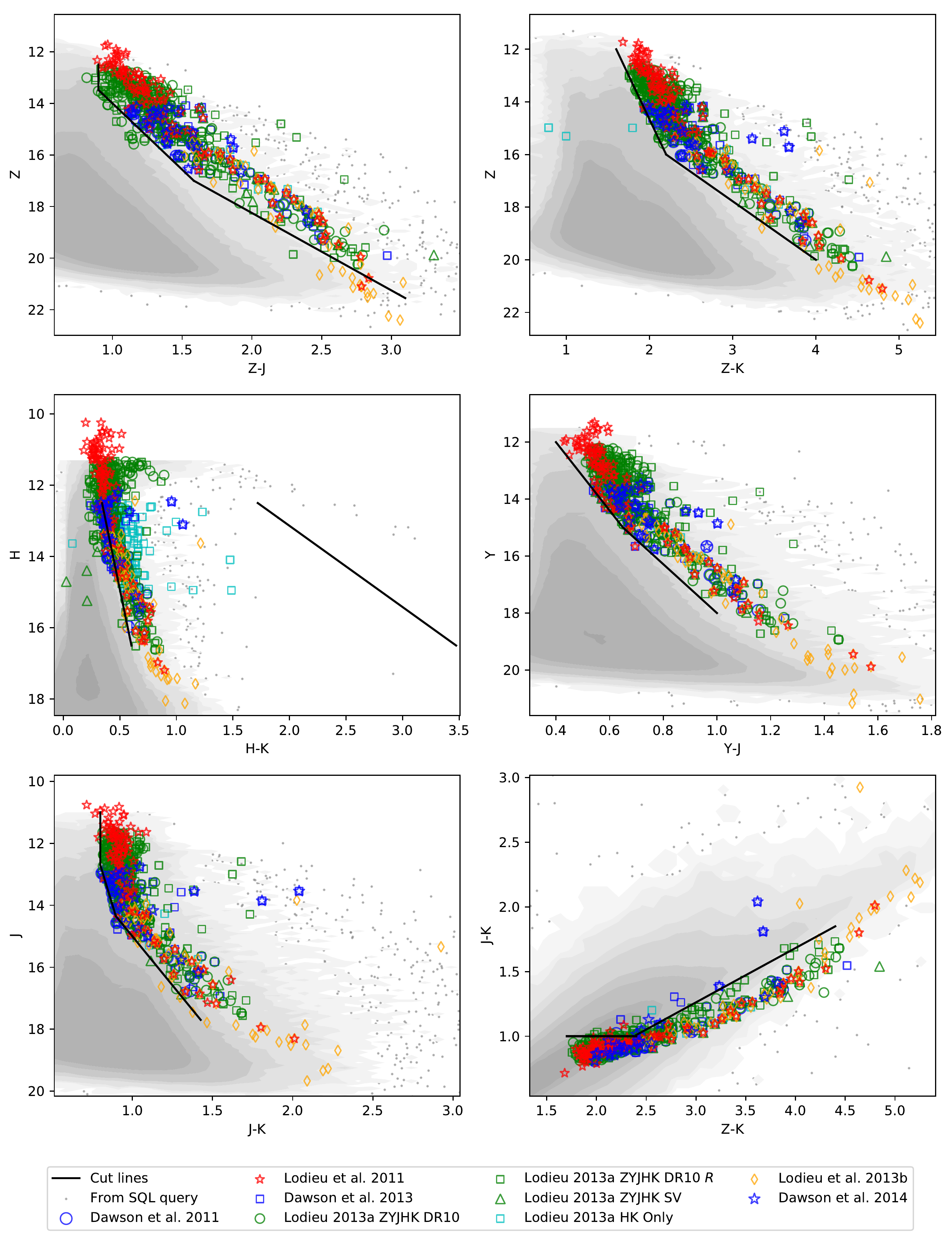}
\caption{Color cuts applied to the samples obtained from the WSA. The gray contours show the data for the `C-ZYJHK DR10' sample before the cuts were applied. Colored symbols show objects from the literature sample. The stars represent sources confirmed spectroscopically (\citealt{Lodieu2011} or \citealt{Dawson2014}). \label{figure:colour_cuts_lit}}
\end{center}
\end{figure*}

\subsection{The L-sample}
\label{section:bds_in_upper_sco:lit_sample}

Although there are many surveys that study \UpperSco, we decided to choose those surveys that identify VLMO members using UKIDSS GCS \citep{Lawrence2007} or similar (\ie VISTA) photometry so that we had sub-samples that had Z, Y, J, H, K photometry ('L-ZYJHK sample'); or had H and K photometry (`L-HK only sample'). We combined data from \citet{Dawson2011}, \citet{Lodieu2011}, \citet{Dawson2013}, \citet{Lodieu2013a}, \citet{Lodieu2013b}, and \citet{Dawson2014} to obtain a sample of 789 unique objects, of which 493 were in the L-ZYJHK sample and 295 were in the L-HK only sample using photometry from both UKIDSS GCS DR10 (henceforth DR10) and the GCS Science verification release \citep[henceforth SV;][]{Dye2006}. Tables \ref{table:samples_by_numbers_l} and \ref{table:samples_by_numbers_c} give the full detail on how many objects were in each source catalog.

\subsection{The C-sample}
\label{section:bds_in_upper_sco:colour_sample}

The data for the C-sample were obtained using the WFCAM Science Archive \citep[WSA][]{Hambly2008}\footnote{WSA available on-line at \url{http://surveys.roe.ac.uk/wsa}.} using SQL queries (see \Appendix\ref{section:appendix_samples}). We followed the initial sampling used by \citet{Lodieu2013a}, using identical bright saturation limits, limiting merged passband selection to 1\arcsec and retaining only point-like, non-duplicated sources. We decided to also follow the sub-sample selection of \citet{Lodieu2013a}, defining samples that had Z, Y, J, H, K photometry ('C-ZYJHK sample'); or as having H and K photometry (`C-HK only sample'). We obtained 2,653,897 sources for the C-ZYJHK sample from DR10, 157,325 sources for the C-ZYJHK sample from SV, and 7,473,530 for the `HK sample' of which 4,814,722 do not have Z, Y and J photometry (\ie the C-HK only sample).

Following \citet{Lodieu2013a} we split the C-ZYJHK sample into a sub-sample affected by reddening and a sub-sample not affected by reddening (henceforth denoted as\Redcut for the reddened sample), and remove those objects with `HK extinction' in the C-HK only sample \citep[see table 1 from ][]{Lodieu2013a}. The C-ZYJHK DR10 sample had 1,722,423 sources flagged as not affected by reddening and 931,474 flagged as being affected by reddening. The C-ZYJHK SV sample had no sources flagged as affected by reddening. The C-HK only sample had 3,652,715 that were not removed due to reddening.

To select VLMOs from the full samples we used the literature sample and the color cuts identified by \citet{Lodieu2013a}. Our final color cuts are nearly identical to \citet{Lodieu2013a} except that we add an additional cut to $H-K$, this was in order to make sure our samples were not affected by the tail of the giant branch (See the $H$ against $H-K$ plot in \Figure\ref{figure:colour_cuts_lit}). The cuts are listed below.

\[
  \text{ZZJ cut} = 
  \begin{cases}
   Z-J > 0.90 & 12.50 < Z < 13.50 \\
   Z < 5.14 (Z-J)+8.86 & 13.50 < Z < 17.00 \\
   Z < 3.00(Z-J) + 12.25 & 17.00 < Z < 21.55 \\
  \end{cases}
\]
\[
  \text{HHK cut} = 
  \begin{cases}
   H < 15.38(H-K)+7.23 & 12.50 < H < 16.50 \\
   H > 2.28(H-K)+8.58 & 12.50 < H < 16.50 \\
  \end{cases}
\]
\[
  \text{JJK cut} = 
  \begin{cases}
   J-K > 0.80 & 11.00 < J < 12.70 \\
   J < 17.00(J-K)-0.90 & 12.70 < J < 14.35 \\
   J < 6.33(J-K) - 8.67 & 14.35 < J < 17.70 \\
  \end{cases}
\]
\[
  \text{ZZK cut} = 
  \begin{cases}
   Z < 6.67(Z-K)+1.33 & 12.00 < Z < 16.00 \\
   Z < 2.22(Z-K)+11.11 & 16.00 < H < 20.00 \\
  \end{cases}
\]
\[
  \text{YYJ cut} = 
  \begin{cases}
   Y < 12.00(Y-J)+7.20 & 12.00 < Y < 15.00 \\
   Y < 8.57(Y-J)+ 9.43 & 15.00 < Y < 18.00 \\
  \end{cases}
\]
\[
  \text{JKZK cut} = 
  \begin{cases}
   (J-K) < 1.00 & 1.70 < (Z-K) < 2.40 \\
   (J-K) < 0.42(Z-K) & 2.40 < (Z-K) < 4.40 \\
  \end{cases}
\]

\vspace{1cm}

We decided to keep two different combinations of these cuts to see the effect they had on the final population selected, the first used all the above cuts (denoted by a \HKcut) and the second used all the color cuts except the `HHK cut'. 

Thus we have samples with `C-ZYJHK DR10', `C-ZYJHK DR10\Redcut', `C-ZYJHK DR10\HKcut', `C-ZYJHK DR10\Redcut\HKcut', `C-ZYJHK SV', `C-ZYJHK SV\HKcut', and `C-HK only' (where the `C' distinguishes the sub-samples from the L-samples described in \Section\ref{section:bds_in_upper_sco:lit_sample}). The number of objects left after the color cuts are shown in \Table\ref{table:colour_cut_numbers} and a full break down of numbers is presented in tables \ref{table:samples_by_numbers_l} and \ref{table:samples_by_numbers_c}.

\begin{table*}
\begin{center}
\caption{The results for the C-samples after the color cuts are applied. \label{table:colour_cut_numbers}}
\begin{tabular}{p{2.75cm}p{2.5cm}p{1cm}p{1cm}p{1cm}p{1cm}p{1cm}p{1cm}p{2.5cm}}
\hline
\hline
Sample & Total before cuts & ZZJ & HHK & JJK & ZZK & YYJ & JKZK & Total after cuts \\ 
\hline
C-ZYJHK DR10 & 1,722,423 & 5,654 & - & 23,134 & 9,977 & 12,122 & 245,810 & 1,305 \\
C-ZYJHK DR10 \Redcut & 931,474 & 1,538 & - & 9,940 & 2,569 & 5,824 & 400,451 & 811 \\
C-ZYJHK SV & 157,325 & 135 & - & 887 & 143 & 206 & 29,723 & 86 \\
\hline
C-ZYJHK DR10 \HKcut & 1,722,423 & 5,654 & 2,359 & 23,134 & 9,977 & 12,122 & 245,810 & 66 \\
C-ZYJHK DR10 \Redcut \HKcut & 931,474 & 1,538 & 318 & 9,940 & 2,569 & 5,824 & 400,451 & 77 \\
C-ZYJHK SV \HKcut & 157,325 & 135 & 71 & 887 & 143 & 206 & 29,723 & 33 \\
\hline
C-HK only & 3,652,715 & - & 1,526 & - & - & - & - & 1,526 \\
\hline
\hline
\end{tabular}

\end{center}
\end{table*}

\subsection{Discussion of sub-samples}
\label{section:bds_in_upper_sco:sub-samples}

In summary we have two samples, the L-sample, constructed directly from the literature and the C-sample, constructed from UKIDSS DR10 and SV SQL Queries and color cuts. The L-sample is split into a `ZYJHK' sample (comprising of objects with Z, Y, J, H and K photometry) and the `HK only' (comprising of those object with only H and K photometry), these are named the `L-ZYJHK' and `L-HK only' samples. The C-sample is also split into a `ZYJHK' and the `HK only' sample (with the same definition), split by the data origin (\ie either UKIDSS GCS DR10 or UKIDSS GCS SV) and by whether we use the `HHK cut' (`HKcut') and whether the area the objects resides is flagged as having reddening (\Redcut). We do this so we can analyze the affect different cuts have on our results. \Table\ref{table:subsample_definition} describes these sample subsets and their differences.

\begin{table*}
\begin{center}
\caption{The definition of the samples \label{table:subsample_definition}}
\begin{tabular}{llp{1cm}p{1cm}p{3cm}p{3cm}p{1.5cm}}
\hline
\hline
Sample & Sub-sample & From DR10 & From SV & Has Z, Y, J, H, and K photometry & Flagged as affected by reddening & \HKcut Cut used \\
\hline
\multirow{2}{*}{l-sample} & L-ZYJHK                         & - 			& - 			& $\checkmark$ 	& - 			& - 			\\
                          & L-HK only 						& - 			& - 			& $\times$		& - 			& - 			\\
\hline
\multirow{7}{*}{c-sample} & C-ZYJHK DR10\noRedcut\HKcut 	& $\checkmark$ 	& $\times$		& $\checkmark$ 	& $\times$		& $\checkmark$ 	\\
                          & C-ZYJHK DR10\Redcut\HKcut 		& $\checkmark$ 	& $\times$		& $\checkmark$ 	& $\checkmark$ 	& $\checkmark$ 	\\
                          & C-ZYJHK SV\HKcut				& $\times$		& $\checkmark$ 	& $\checkmark$ 	& $\times$		& $\checkmark$ 	\\
                          & C-ZYJHK DR10\noRedcut\noHKcut 	& $\checkmark$ 	& $\times$		& $\checkmark$ 	& $\times$		& $\times$		\\
                          & C-ZYJHK DR10\Redcut\noHKcut 	& $\checkmark$ 	& $\times$		& $\checkmark$ 	& $\checkmark$ 	& $\times$		\\
                          & C-ZYJHK SV 						& $\times$		& $\checkmark$ 	& $\checkmark$ 	& $\times$		& $\times$		\\
                          & C-HK only 						& $\checkmark$ 	& $\times$		& $\times$		& $\times$		& $\checkmark$ 	\\
\hline
\hline
\end{tabular}

\end{center}
\end{table*}

Each of our sub-samples has specific properties due to the imposed selection criteria. The C-HK only sample is going to be the least well defined sample due to the lack of Z, Y and J photometry and therefore lack of all the color cuts except the \HKcut cut. The \Redcut samples (\ie C-ZYJHK DR10 \Redcut and C-ZYJHK DR10\Redcut\HKcut) are expected to be more contaminated due to the increased reddening those source experience (\ie reddened objects will be scattered across the color cuts). The SV samples (\ie C-ZYJHK SV and C-ZYJHK SV\HKcut) rely on science verification photometry and will be less complete than the DR10 samples. The SV samples also cover a slightly different spatial location than the DR10 samples and thus may have slight differences in age (see the age gradient from figure 9 of \citealt{Pecaut2016}). The L-sample consists of some higher mass objects (due to the lack of, for example, brightness cuts), however due to some of these objects being spectroscopically confirmed we do not further reduce the L-sample and use it for comparative purposes only. 

Our C-samples are much improved over the previously existing L-sample (due to our use of the best available color cuts and the better proper motion cuts (see \Section\ref{section:pm_analysis}). The addition of the \HKcut cut to the C-ZYJHK samples add a brightness cut which essentially acts as a mass cut ($\lesssim 0.2\Msun$ see effects in \Section\ref{section:properties:mass_function} and \ref{section:properties:discs}). This cut has the effect of avoiding contamination from red giants at the bright end of the magnitude distribution. Therefore, the C-ZYJHK DR10\noRedcut\HKcut should be the least contaminated sample among our sub-samples, but also the most restrictive. On the other hand, the C-ZYJHK DR10 \noRedcut\noHKcut should be the most complete. Throughout this paper we do all analysis on all samples to check how much of an effect selection has on any results we obtain.

\section{Proper motion analysis}
\label{section:pm_analysis}

\begin{figure*}
\begin{center}
\includegraphics[width=.8\textwidth]{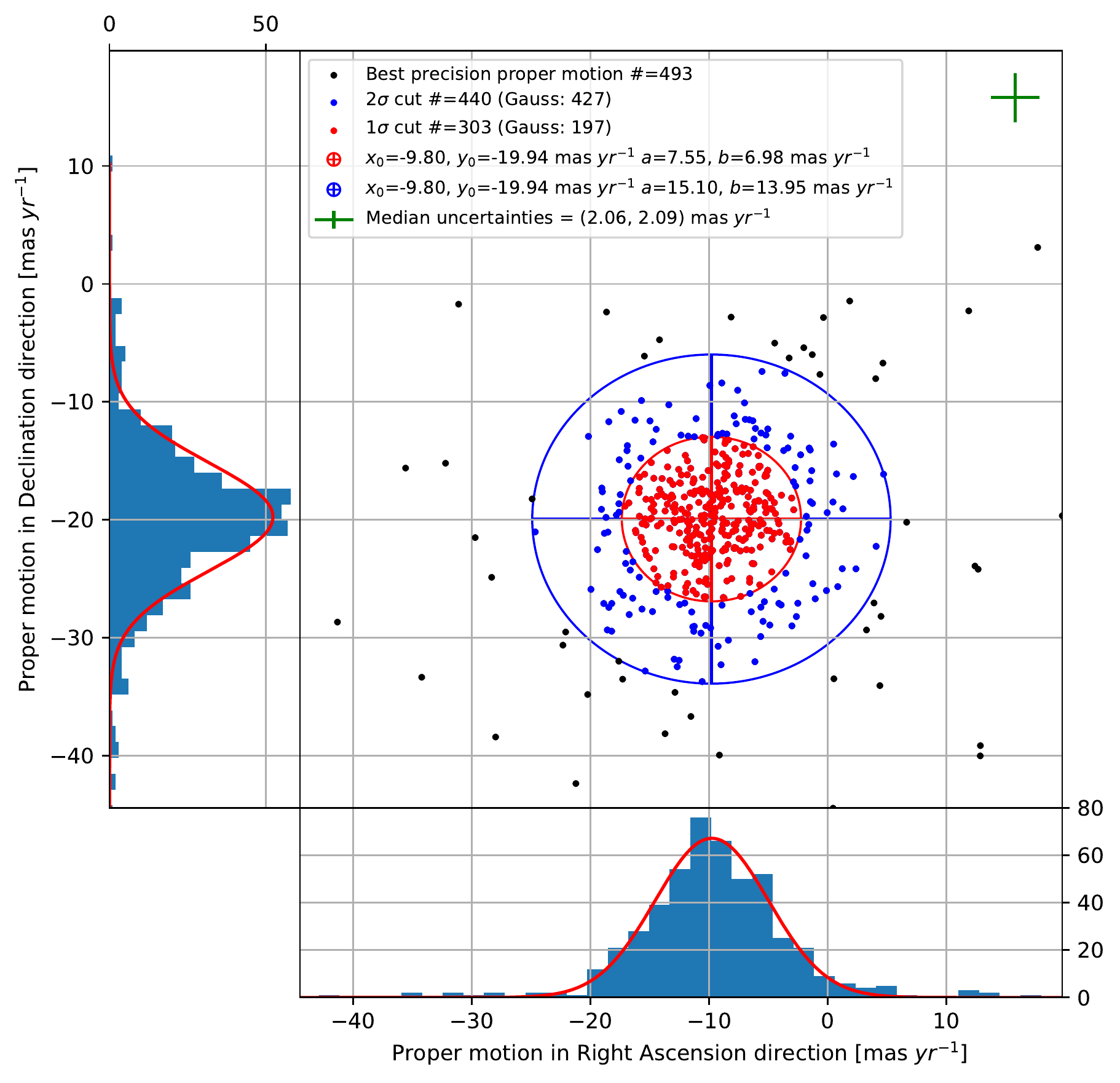}
\caption{Proper motion vector diagram showing the one and two sigma ellipses used to define \UpperSco membership for the L-ZYJHK sample (for the best precision proper motions). The numbers of objects found were compared to a two-dimensional Gaussian distribution of equal center and covariance. Median uncertainties are shown in the upper right corner. \label{figure:pm_vector_diagram}}
\end{center}
\end{figure*}

As with many previous moving group membership surveys (\eg \citealt{Dawson2011,Dawson2013}; \citealt{Lodieu2013a,Lodieu2013b}) we define \UpperSco membership as having a proper motion in both the Right Ascension and Declination directions consistent with that of \UpperSco. Since the release of the Gaia DR1 there have been new catalogs generated using the Gaia DR1 positions for objects without any Gaia proper motions \citep[\ie the Gaia DR1 secondary catalog of $\sim$1.1 billion sources,][]{Gaia2016b}. Three of the largest proper motion catalogs currently using Gaia DR1 positions (and overlapping on-sky with \UpperSco) are the {\itshape Hot Stuff for One Year catalog,} \citep[HSOY; containing $\sim$583 million stars][]{Altmann2017}, the {\itshape Gaia-PS1-SDSS proper motion catalog} \citep[GPS1; containing $\sim$350 million stars,][]{Tian2017}, and the {\itshape US Naval Observatory CCD astrograph catalog 5} \citep[UCAC5; containing $\sim$107 million stars,][]{Zacharias2017}. We cross-matched (selecting the closest source within a 3\arcsec matching radius) all sub-samples (belonging to both the L-sample and the C-sample) with HSOY, GPS1 \citep[also giving us access to the {\itshape Pan-STARRS1 proper motions}; PS1][]{Chambers2011}, UCAC5, as well as the PPMXL catalog \citep[containing $\sim$900 million sources]{Roeser2010}, and the proper motions associated with each source from UKIDSS GCS DR10. From these proper motions the most precise total proper motion was selected for each object (where total proper motion and associated uncertainty are defined in \Equation\ref{equation:total_proper_motion}). 

\begin{equation}
\label{equation:total_proper_motion}
\mu_{Total} = \sqrt{\mu_{\alpha}^2 + \mu_{\delta}^2}\hspace{2cm}
\sigma_{\mu_{Total}} = \left|{\frac{1}{\mu_{Total}}}\right|\sqrt{(\sigma_{\mu_{\alpha}}\mu_{\alpha})^2 + (\sigma_{\mu_{\delta}}\mu_{\delta})^2}
\end{equation}

\noindent where $\mu_{\alpha}$ is the proper motion component in the Right Ascension direction ($cos(\delta)$) and $\mu_{\delta}$ is the proper motion component in the Declination direction.

We decided to exclude PPMXL proper motions as no sources with only PPMXL proper motions had uncertainties better than $\sim$10 \masyr. The stars that had a suitable proper motion measurement 716 out of 789 for the L-sample, all 2259 C-ZYJHK DR10 objects, 68 out of 86 C-ZYJHK SV objects, all 17 C-ZYJHK SV\HKcut, 1,519 out of 1,526 C-HK only sample (see tables \ref{table:samples_by_numbers_l} and \ref{table:samples_by_numbers_c} for a full break down of numbers).

\begin{table}
\begin{center}
\caption{The result of the \UpperSco membership selection. \label{table:pm_breakdown}}
\begin{tabular}{lcc}
\hline
\hline
Sample & Total before cuts & Total after pm cuts \\ 
\hline
L-ZYJHk & 453 & 415 \\
L-HK & 241 & 175 \\

\hline
C-ZYJHK DR10 & 1,305 & 171 \\
C-ZYJHK DR10 \Redcut & 881 & 224 \\
C-ZYJHK SV & 68 & 58 \\
\hline
C-ZYJHK DR10 \HKcut & 66 & 49 \\
C-ZYJHK DR10 \Redcut \HKcut & 77 & 68 \\
C-ZYJHK SV \HKcut & 17 & 14 \\
\hline
C-HK only & 1,519 & 346 \\
\hline
\hline
\end{tabular}
\end{center}
\end{table}

Our uncertainties in proper motion are sufficiently small that we do not need to select members based on a proper motion uncertainty circle (this is in contrast to previous studies where large uncertainties dominate the velocity dispersion of \UpperSco). However, since the proper motion of \UpperSco is very small, we decided to use the L-sample to define a two-sigma membership ellipse for \UpperSco (such that we avoid an overlap with $(\mu_{\alpha},\mu_{\delta})=(0.0, \,0.0)$ \masyr).  We compared the median and standard deviations for the L-ZYJHK, L-HK only and the combined sample. The L-ZYJHK sample was found to have a center of $(\mu_{\alpha}, \mu_{\delta})$=(-9.80, -19.94) \masyr, with standard deviations of $(\mu_{\alpha}, \mu_{\delta})$=(7.51, 6.96) \masyr. The L-HK only sample was found to have a center of $(\mu_{\alpha}, \mu_{\delta})$=(-7.78, -18.39) \masyr, with standard deviations of $(\mu_{\alpha}, \mu_{\delta})$=(7.82, 9.14) \masyr. The combined L-sample was found to have a center of $(\mu_{\alpha}, \mu_{\delta})$=(-9.04, -19.46) \masyr, with standard deviations of $(\mu_{\alpha}, \mu_{\delta})$=(7.88, 8.08) \masyr. 

We thus chose to define candidate members of \UpperSco as those within an ellipse of center $(\mu_{\alpha}, \mu_{\delta})$=(-9.80, -19.94) and x and y radii of $(\mu_{\alpha}, \mu_{\delta})$=(15.10, 13.95) \masyr (defined from the L-ZYJHK distribution). This was then used to select members from the L-sample and C-sample. We keep 415/453 and 175/241 of those objects from the L-ZYJHK and L-HK only samples respectively. For the C-ZYJHK DR10\Redcut sample we kept in 224/881 candidates, and 68/77 C-ZYJHK DR10\Redcut\HKcut candidates. For the C-ZYJHK\noRedcut\, sample we identified 171/1,305 candidates, and for the C-ZYJHK\noRedcut\HKcut sample we kept 49/66 candidates. For the C-ZYJHK SV sample we identify 58/68 as \UpperSco candidates, and 14/17 for the C-ZYJHK SV\HKcut sample. The C-HK only sample resulted in 346/1,519 candidates being identified. These numbers are summarized in \Table\ref{table:pm_breakdown} and tables \ref{table:samples_by_numbers_l} and \ref{table:samples_by_numbers_c} have a full break down of numbers. \Figure\ref{figure:pm_vector_diagram} shows the L-ZYJHK sample used to selected candidates from both the L-samples and the C-samples.

The addition of a proper motion cut does confirm the characterization of the samples given in \Section\ref{section:bds_in_upper_sco:sub-samples}. Specifically, the C-HK sample without ZYJ photometry is heavily contaminated, as expected. The same applies to the C-ZYJHK DR10 and C-ZYJHK DR10 \Redcut samples. We expect most of the contamination in these samples to be at the bright magnitude end, \ie at high masses in the VLM domain, because in this regime the population of young \UpperSco members is not well separated from the background population in color-magnitude diagrams. As mentioned above, the cleanest sample in our list are the ones with the \HKcut cut, which removes objects at the bright end.

\section{Properties of VLMOs in Upper Scorpius}
\label{section:properties}

With an estimated age of 10 \Myr (with a spread between 7 and 15 \Myr) and assuming a distance of $\sim$145 pc (with a spread of $\pm$13 pc, \Section\ref{section:upper_sco}) it is possible to estimate mass and luminosity by fitting the photometry to theoretical isochrones. We use the 8, 10 and 15 \Myr \citet{Baraffe2015} isochrones (BHAC15) to give a lower, median and upper bound to each of our objects with UKIDSS Z, Y, J, H and K photometry (\ie we only fit sources which have all five photometric magnitudes), we choose 8\Myr as the lower bound as the 7\Myr isochrone is not computed for BHAC15. In this section we describe the fitting process and use these, with {\itshape Wide-Field Infrared Survey Explorer} \citep[WISE,][]{Wright2010} data to infer a disc fraction, analyze the mass distributions and explore the proper motion distribution of our candidates.

\begin{figure*}
\begin{center}
\includegraphics[trim={0 0 0 2.25cm}, clip, width=.8\textwidth]{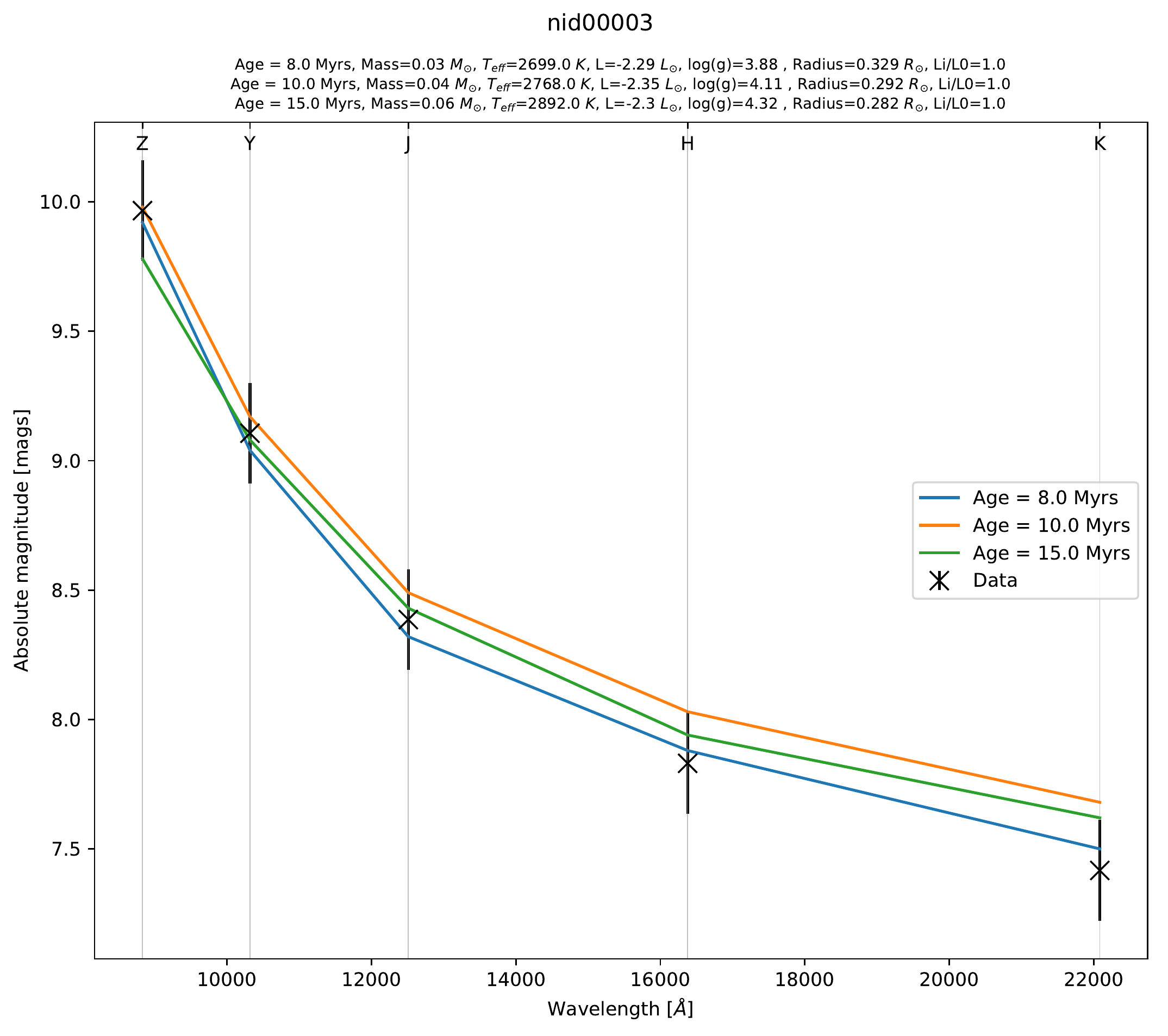}
\caption{Example isochronal fit for L-sample object UGCS J161625.98-211222.9. Fit gives a mass of 0.04$^{+0.02}_{-0.01}$\Msun. \label{figure:example_isochrone_fit}}
\end{center}
\end{figure*}

\begin{figure*}
\begin{center}
\includegraphics[width=\textwidth]{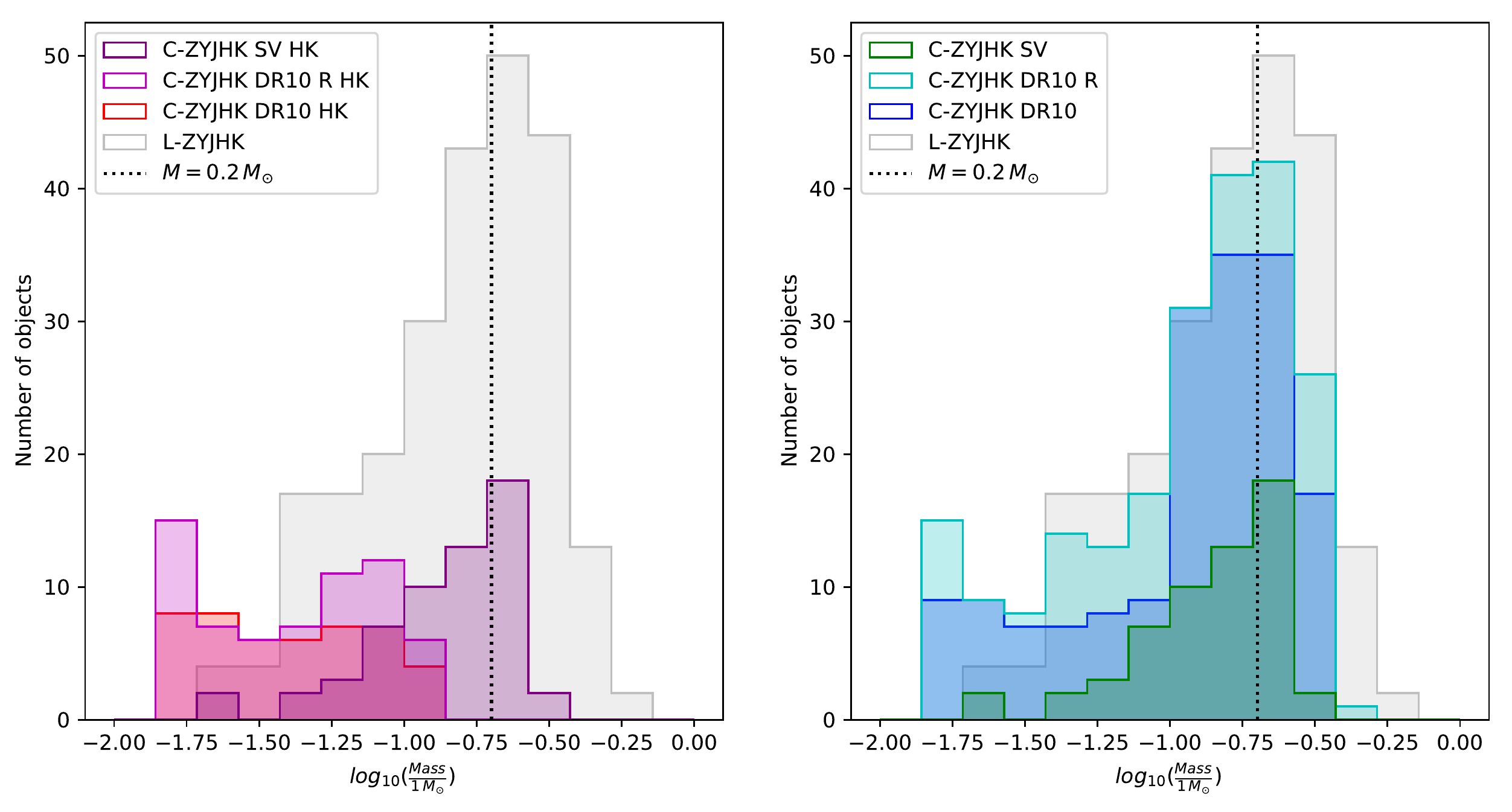}
\caption{Log mass histogram for the C-sample using the \HKcut cut (left) as compared to cases without (right). The \HKcut cut effectively constrains the mass of the objects to $\sim$0.2\Msun (black vertical line) whereas without the \HKcut cut the masses extend to higher mass objects. Over-plotted, in both sub-plots for reference, is the L-sample (with no \HKcut cut applied).  \label{figure:mass_vs_zj}}
\end{center}
\end{figure*}

\begin{figure*}
\begin{center}
\includegraphics[width=.8\textwidth]{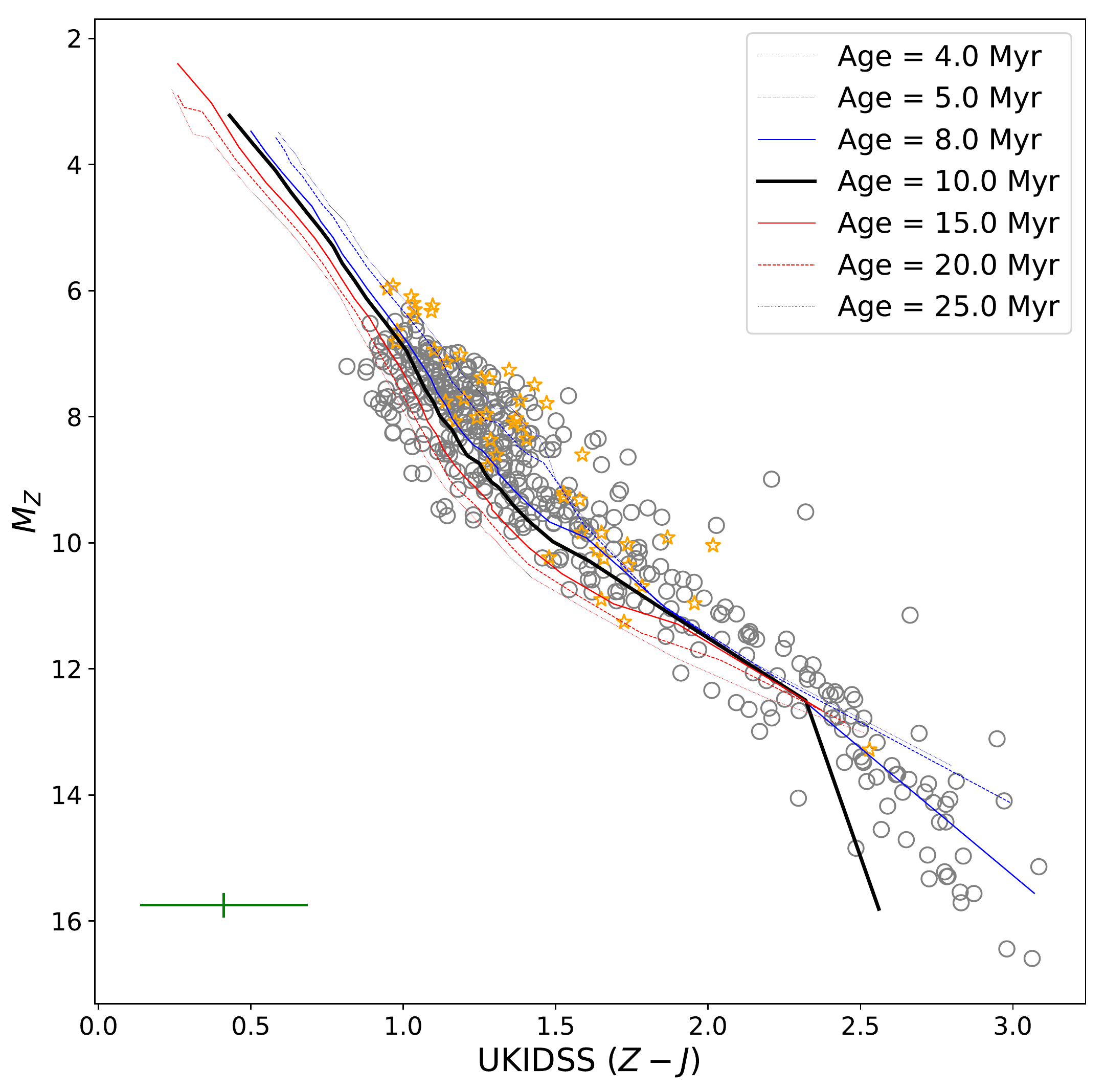}
\caption{Absolute Z magnitude against (Z-J) color (Hertzsprung-Russell diagram) for the L-sample. Objects with discs are marked with an orange star. Median uncertainties are shown with the green cross. The distribution is consistent with an average age of 10 \Myr and a spread from 8 to 15 \Myr. \label{figure:HR-lsample}}
\end{center}
\end{figure*}

\subsection{Isochronal fitting}
\label{section:properties:isofit}

Using the 8, 10 and 15 \Myr BHAC15 isochrones for UKIDSS we used chi-squared minimization (using the apparent UKIDSS magnitudes converted to absolute magnitudes using a distance of 146 pc, see \Section\ref{section:upper_sco}) to select the best fit model for a lower, median and upper bounding model. The nature of the BHAC15 isochrones means for a given set of photometry (\i.e. Z,Y,J,H and K) and age we get a mass estimate for each object (with an associated luminosity, effective temperature, \Teff, radius or surface gravity, $log(g)$ for each mass estimate).

The mass estimate attached to each age (8, 10 and 15 \Myr) were then combined, giving an expected value, an upper and a lower uncertainty (described in \Equation\ref{equation:isochrone_bounds}).

\begin{equation}
\label{equation:isochrone_bounds}
\begin{split}
\mathcal{M} & = (\mathcal{M}_{\text{median}})^{+\sigma_{\mathcal{M}_{upper}}}_{-\sigma_{\mathcal{M}_{lower}}} \\
\mathcal{M}_{\text{median}} & = \text{median fit} \\
\mathcal{M}_{\text{lower}} & = \text{lower fit} \\
\mathcal{M}_{\text{upper}} & = \text{upper fit} \\
\sigma_{\mathcal{M}_{lower}} & = 
                        \begin{cases} 
                        \mathcal{M}_{\text{upper}} - \mathcal{M}_{\text{median}} & \text{if } \mathcal{M}_{\text{lower}}=0 \text{ and } \mathcal{M}_{\text{upper}}\ne0 \\
                        \mathcal{M}_{\text{upper}} - \mathcal{M}_{\text{median}} & \text{if } \mathcal{M}_{\text{lower}}<0 \text{ and } \mathcal{M}_{\text{upper}}<0 \\ 
                        \mathcal{M}_{\text{median}} & \text{if } \mathcal{M}_{\text{lower}}=\mathcal{M}_{\text{upper}}=0 \\
                        \mathcal{M}_{\text{median}} - \mathcal{M}_{\text{lower}} & \text{else-wise}
                        \end{cases} \\
\sigma_{\mathcal{M}_{upper}} & = 
                        \begin{cases} 
                        \mathcal{M}_{\text{median}} - \mathcal{M}_{\text{lower}} & \text{if } \mathcal{M}_{\text{lower}}=0 \text{ and } \mathcal{M}_{\text{upper}}\ne0 \\
                        \mathcal{M}_{\text{median}} - \mathcal{M}_{\text{lower}} & \text{if } \mathcal{M}_{\text{lower}}<0 \text{ and } \mathcal{M}_{\text{upper}}<0 \\ 
                        \mathcal{M}_{\text{median}} & \text{if } \mathcal{M}_{\text{lower}}=\mathcal{M}_{\text{upper}}=0 \\
                        \mathcal{M}_{\text{upper}} - \mathcal{M}_{\text{median}} & \text{else-wise}
                        \end{cases} \\
\end{split}
\end{equation}

\noindent where $\mathcal{M}$ is the mass estimate associated with the best fit (lower, median and upper bounding) model. This gave us appropriate uncertainties for the estimated mass based on the spread in ages found for \UpperSco (\Section\ref{section:upper_sco}). We do not interpolate between the models and these estimated masses assume that all objects have an age between 8 and 15 \Myr with a median of 10 \Myr. An example fit can be seen in \Figure\ref{figure:example_isochrone_fit} for L-sample object UGCS J161625.98-211222.9. 

The estimated mass distributions for the C-sample using the \HKcut cut as compared to cases without can be seen in \Figure\ref{figure:mass_vs_zj} (with the L-sample over-plotted in both cases for comparison). From \Figure\ref{figure:mass_vs_zj} the differences between the sub-samples becomes clear. The L-samples contain a significant number of higher-mass stars. The application of the \HKcut cut (left panel compared to right panel) shows that this cut effectively removes objects of mass $\lesssim 0.2\Msun$ and hence avoids contamination. The Hertzsprung-Russell diagram for the L-sample is shown in \Figure\ref{figure:HR-lsample}. The distribution of colors seems consistent with a typical age of 10 \Myr. 

Defining brown dwarfs to have a mass less than 0.075\Msun we calculated the number of our objects that are likely brown dwarfs (with the uncertainty coming from those that overlap in mass due to their mass estimate uncertainty). The L-ZYJHK sample was found to have 152$\pm$38 out of the 415 objects as likely brown dwarfs, the C-ZYJHK DR10\HKcut, C-ZYJHK DR10 \Redcut\HKcut and C-ZYJHK SV\HKcut samples were found to have 42$\pm$11, 53$\pm$10 and 10$\pm$6 respectively, and the C-ZYJHK DR10, C-ZYJHK DR10 \Redcut and C-ZYJHK SV samples were found to have 48$\pm$13, 67$\pm$13 and 10$\pm$6 respectively. These numbers are presented in \Table\ref{table:number_possible_bds}. We note that the number of brown dwarfs are quite similar in the samples with and without the \HKcut cut; in this mass domain the samples should avoid most contamination.

\begin{table*}
\begin{center}
\caption{Numbers of possible brown dwarfs (Mass less than 0.075\Msun). \label{table:number_possible_bds}}
\begin{tabular}{p{6cm}p{4cm}p{1.5cm}}
\hline
\hline
Sample & Total objects in sample & Likely brown dwarfs \\
\hline
L-ZYJHK 						 & 415 & 152$\pm$38 \\
\hline
C-ZYJHK DR10\noRedcut\HKcut		 &  49 &  42$\pm$11 \\
C-ZYJHK DR10\Redcut\HKcut		 &  68 &  53$\pm$10 \\
C-ZYJHK SV\HKcut				 &  14 &  10$\pm$6  \\
\hline
C-ZYJHK DR10\noRedcut			 & 171 &  48$\pm$13 \\
C-ZYJHK DR10\Redcut				 & 224 &  67$\pm$13 \\
C-ZYJHK SV						 &  58 &  10$\pm$6  \\
\hline
\hline
\end{tabular}\end{center}
\end{table*}

We were concerned that any objects with discs (see \Section\ref{section:properties:discs}){} may, due to their young age, have H and K photometry that is not well represented by one of the BHAC15 isochrones (\ie there should be an additional component added to the flux due to the presence of a disc). For this reason we also fitted the masses using only ZYJ and ZYJH and flagged any objects which were identified as having possible discs (from \Section\ref{section:properties:discs}). This led to having three mass estimates for each object and thus we were able to see any differences due to `bad' H and K photometry. After comparing objects with possible discs to those without discs no discernible difference was seen, thus the mass estimates were not affected by those objects having discs. Most objects had a maximum variation (due to using different sets of photometry) of 0.01$\Msun$ thus we decided to retain our mass estimates using all five bands.

Our chi-squared minimization does not take into account the uncertainties in magnitude (shown on \Figure\ref{figure:example_isochrone_fit}) and thus it does not take into account any uncertainty due to distance or spread in the association. Note that the spread in the association is significantly larger than the distance uncertainty on \UpperSco (145$\pm$2 pc), \ie $\pm$13 pc from \Section\ref{section:upper_sco}, thus an additional uncertainty on the mass estimates will be introduced (this will be solved with later Gaia data releases giving distances to individual objects).

\begin{figure*}
\begin{center}
\includegraphics[width=\textwidth]{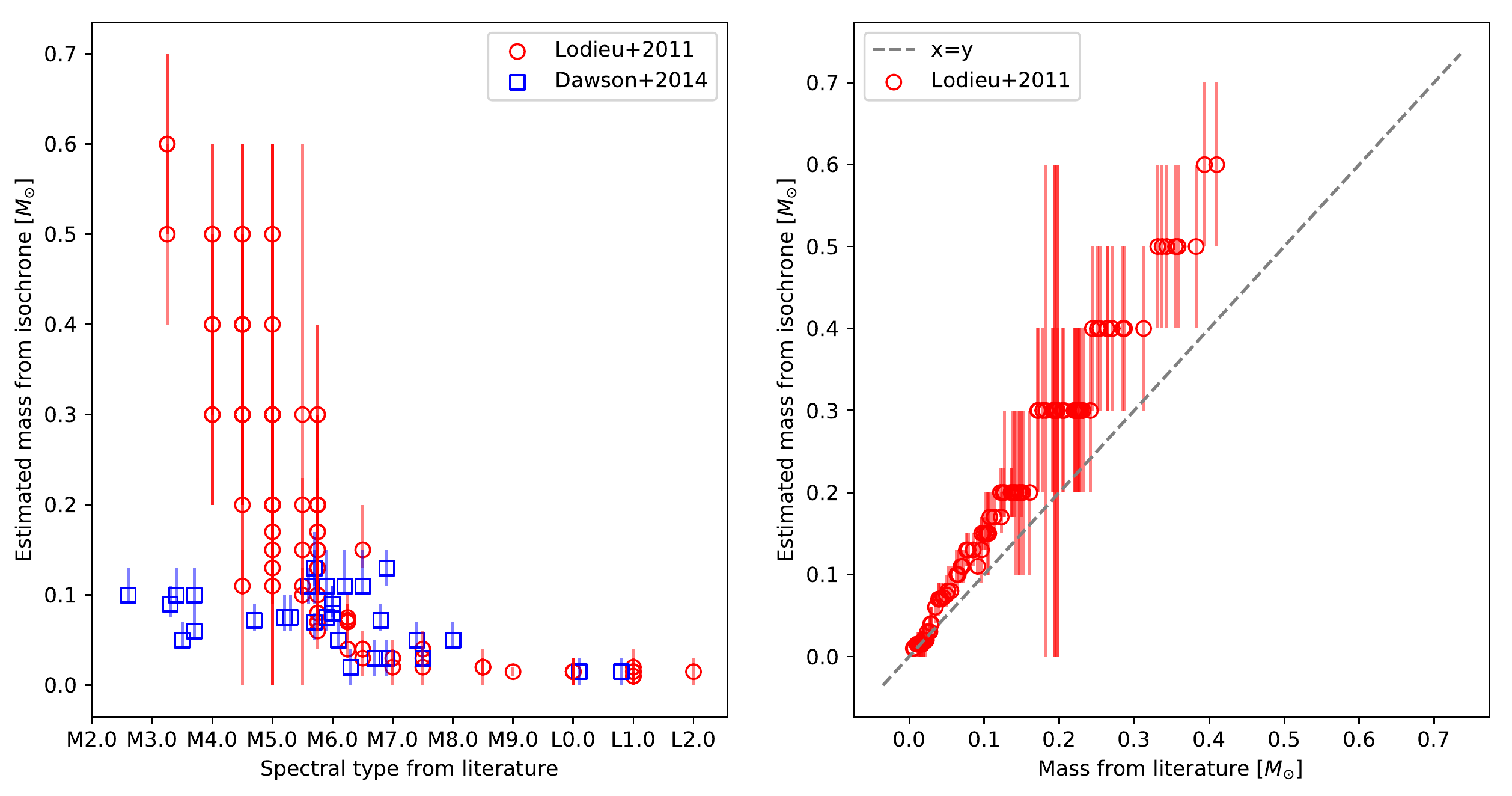}
\caption{The comparison between our mass estimates from the isochrones to those objects in \UpperSco with spectral type and masses from the literature (members with spectra) from \protect\citet{Lodieu2011} and \protect\citet{Dawson2014}. \label{figure:mass_spt_iso_comp}}
\end{center}
\end{figure*}

As a further sanity check for our mass estimates we compared our results to the \citet{Lodieu2011} and \citet{Dawson2014} samples which both have spectroscopically confirmed members of \UpperSco. We compare the spectral types from the literature to the mass estimates from the isochrones and compare the mass estimates from the literature to the mass estimates from the isochrone fits (see \Figure\ref{figure:mass_spt_iso_comp}). The figures show the expected trends and broad agreement, but there are also clear discrepancies. In the left panel of \Figure\ref{figure:mass_spt_iso_comp} some objects (with very low masses) have surprisingly early spectral types, \citet{Dawson2014} concluded that some of these objects might be further away than the rest of the objects identified as being part of \UpperSco. For these objects our mass estimate will be underestimated. In the right panel of \Figure\ref{figure:mass_spt_iso_comp} we find that our mass estimates are systematically higher (by $\sim1\sigma$) compared to \citet{Lodieu2011}. That study derives masses by comparing bolometric magnitudes (derived from J-band) with NextGen/DUST models (\citealt{Baraffe1998} and \citealt{Chabrier2000} respectively), using a distance consistent with ours, but assume an age of 5\,Myr (priv. comm. Lodieu 2017). Between 5 and 10\,Myr, VLMOs drop in luminosity, i.e. assuming a younger age leads to lower mass estimates.

\begin{figure*}
\begin{center}
\includegraphics[width=\textwidth]{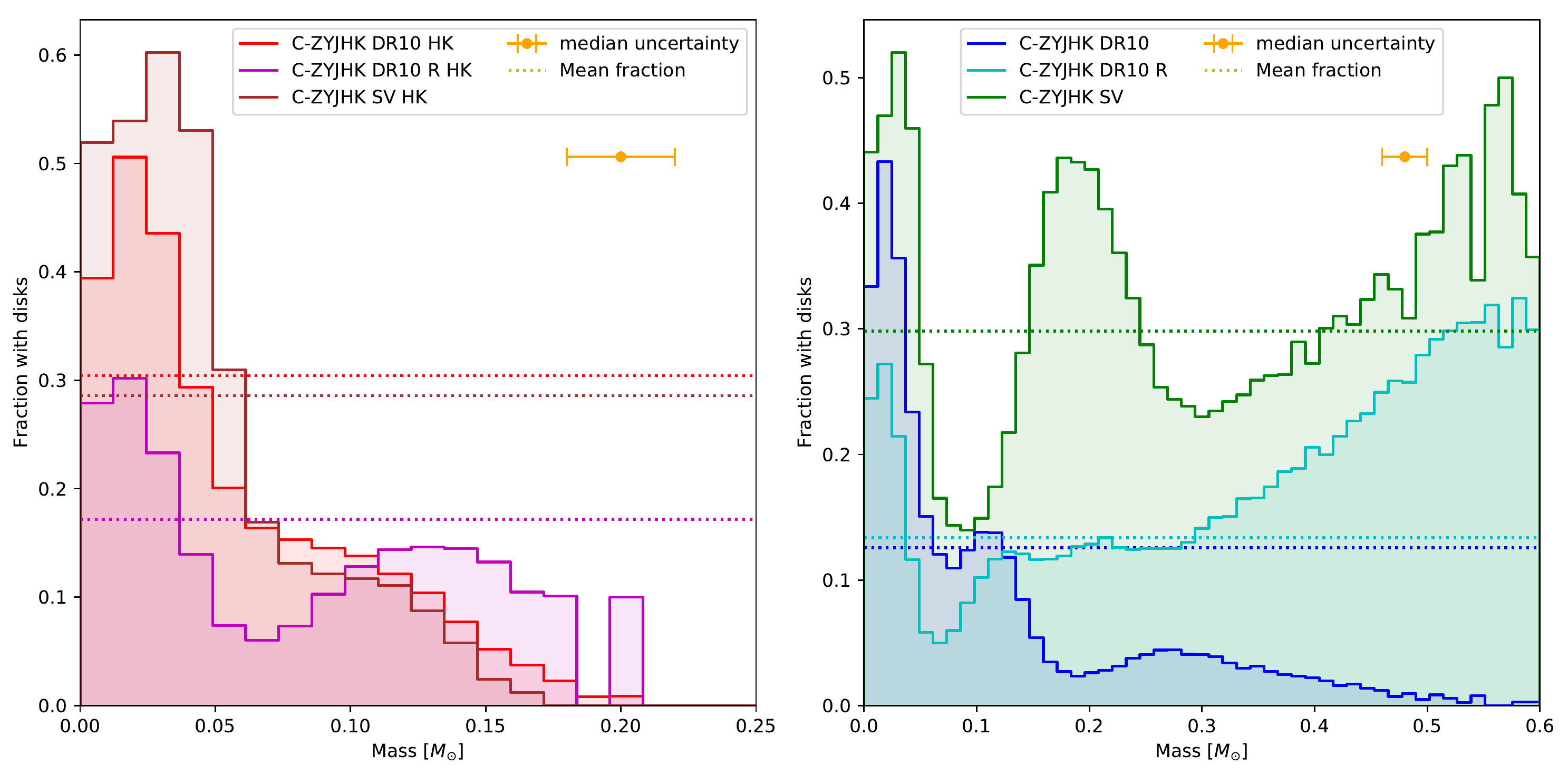}
\caption{Disc fraction as a function of estimated mass (from isochrones). All samples were chosen to have 50 bins ranging from 0.0 to 0.6\Msun (bin sizes were chosen to represent, approximately, the median uncertainty in mass estimates) and used 10,000 samples in our Monte-Carlo analysis (see \Section\ref{section:properties:discs}) for those objects with mass estimates and with WISE photometry. Those flagged with discs were separated and fractions of object per mass bin were calculated. For those samples with the \HKcut cut (left panel) we focus on the region from 0 -- 0.25\Msun level as no objects in the C-ZYJHK samples had larger masses. \label{figure:disc_as_function_of_mass}}
\end{center}
\end{figure*}

\begin{figure*}
\begin{center}
\includegraphics[width=\textwidth]{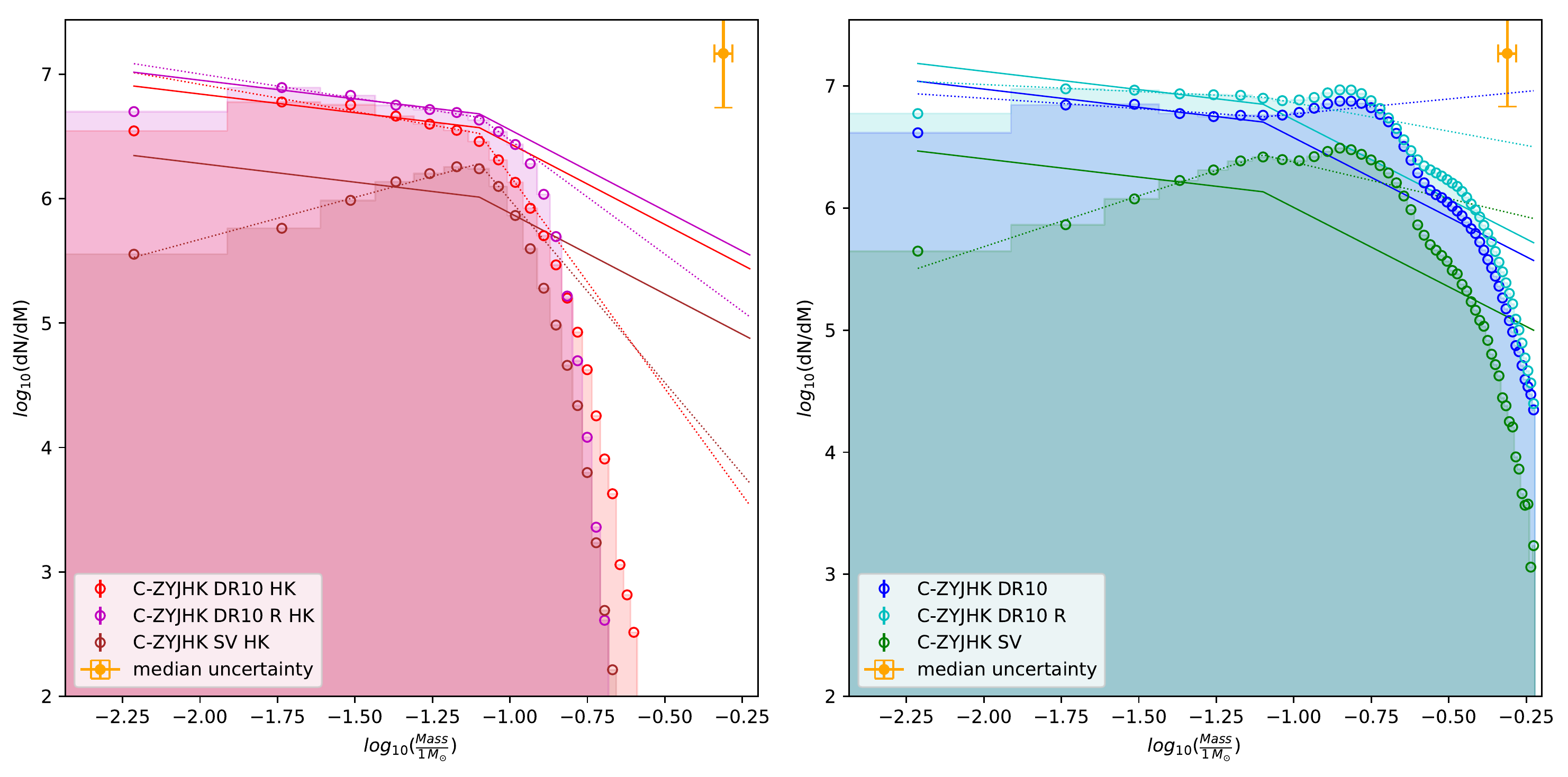}
\caption{Mass functions (from isochrones). All samples were chosen to have 50 bins ranging from 0.0 to 0.6\Msun (bin sizes were chosen to represent, approximately, the median uncertainty in mass estimates) and used 10,000 samples in our Monte-Carlo analysis (see \Section\ref{section:properties:discs}) for those objects with mass estimates. The dashed lines show the best fits to a mass function ($\alpha_1$ for $0.08<M<0.5$\Msun, $\alpha_2$ for $M<$0.08\Msun), where $\alpha$ is allowed to vary and the solid lines show the Kroupa mass function ($\alpha = 1.3$ for $0.08<M<0.5$\Msun, $\alpha=0.3$ for $M<$0.08\Msun). All fits are scaled arbitrarily. \label{figure:mass_function}}
\end{center}
\end{figure*}

\subsection{Disc fraction as a function of mass}
\label{section:properties:discs}

Using the \WaWb color excess cut and W3 excess cut from \citet[][shown in \Equation\ref{equation:disc_cuts}]{Dawson2013} we were able to identify possible discs in our candidate members. 

\begin{equation}
\label{equation:disc_cuts}
\begin{split}
\text{W1W2 Disc} &= J < 60(W1-W2)-9 \\
\text{W3 Disc} &= SNR_{W3} > 5 \text{ and } W3 < 10 \\
\text{Disc} &= (\text{W1W2 Disc}) \text{ or } (\text{W3 Disc}) \\
\end{split}
\end{equation}

For the L-ZYJHK sample we found 47/244 of the \UpperSco members with mass estimates had discs (19.3$\pm$2.5\%). For the C-ZYJHK DR10\noRedcut\HKcut, C-ZYJHK DR10\Redcut\HKcut and C-ZYJHK SV\HKcut samples we found 14/46, 11/64 and 4/14 respectively (30.4$\pm$6.8\%, 17.2$\pm$4.7\% and 28.6$\pm$12.1\%). For the C-ZYJHK DR10\noRedcut\noHKcut, C-ZYJHK DR10\Redcut\noHKcut and C-ZYJHK SV\noHKcut\,samples we found 21/167, 29/217 and 17/57 respectively (12.6$\pm$2.6\%, 13.4$\pm$2.3\% and 29.8$\pm$6.1\%), all fractions are presented in \Table\ref{table:disc_fractions}. Combined these give a weighted mean disc fraction of 16.8$\pm$1.7\%. All uncertainties are calculated as one-sigma uncertainties assuming binomial statistics (see tables \ref{table:samples_by_numbers_l} and \ref{table:samples_by_numbers_c} for a full break down of numbers). This value is broadly consistent with previously published disc fractions for VLM stars and brown dwarfs in this region \citep{Jayawardhana2003,Scholz2007}.

\begin{table*}
\begin{center}
\caption{The disc fractions found for the \UpperSco objects with WISE photometry. \label{table:disc_fractions}}
\begin{tabular}{p{5cm}p{3cm}p{3cm}p{3cm}}
\hline
\hline
Sample & Number of objects with discs & Number of objects in sample & Disc fraction \\
\hline
L-ZYJHK 						& 47 & 244 & 19.3$\pm$2.5\% \\
\hline
C-ZYJHK DR10\noRedcut\HKcut		& 14 &  46 & 30.4$\pm$6.8\% \\
C-ZYJHK DR10\Redcut\HKcut		& 11 &  64 & 17.2$\pm$4.7\% \\
C-ZYJHK SV\HKcut				&  4 &  14 & 28.6$\pm$12.1\% \\
\hline
C-ZYJHK DR10\noRedcut			& 21 & 167 & 12.6$\pm$2.6\% \\
C-ZYJHK DR10\Redcut				& 29 & 217 & 13.4$\pm$2.3\% \\
C-ZYJHK SV						& 17 &  57 & 29.8$\pm$6.1\% \\
\hline
\hline
\end{tabular}\end{center}
\end{table*}

We then used the mass estimates to plot disc fraction as a function of mass, where we account for uncertainties in the mass using a Monte-Carlo approach to draw samples from a Gaussian distribution for each object and then bin up the total samples for candidates, see \Figure\ref{figure:disc_as_function_of_mass}. For each object our Monte-Carlo approach draws a large number of samples from a Gaussian mass distribution (using the full-width-half-maximum as the larger of the lower and upper uncertainty band). Thus our binning process takes into account the uncertainties in mass estimates and shows the distribution as if there were a larger number of objects (note all disc fractions are carefully extrapolated in this process).

\begin{figure*}
\begin{center}
\includegraphics[width=.5\textwidth]{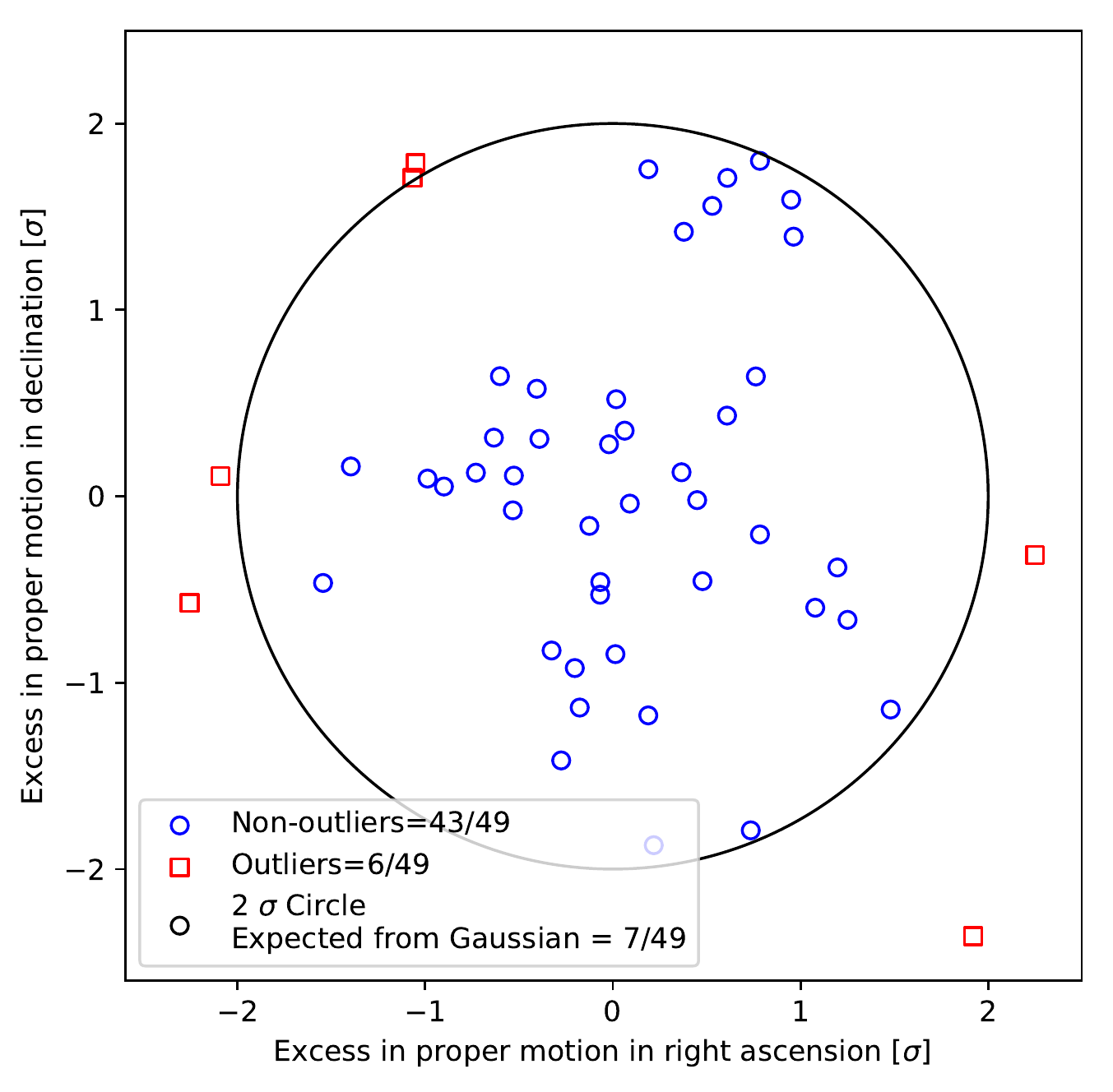}
\caption{Proper motion excess diagrams for the C-ZYJHK DR10\HKcut sample. Excess is defined in \Equation\ref{equation:excess_pm}. \label{figure:pm_outliers}}
\end{center}
\end{figure*}

The disc fractions derived in that manner give two important insights. One is that the choice of the sample has a non-negligible effect on the outcome. This is particularly apparent from the right panel in \Figure\ref{figure:disc_as_function_of_mass}, which shows the sample without the \HKcut cut. Based on our assessment of contamination of these samples (see \Section\ref{section:bds_in_upper_sco:sub-samples} and \Section\ref{section:pm_analysis} the disc fractions derived from these samples have to be treated with caution. Contamination by background red giants might increase the fraction of objects with infrared excess in these samples, while contamination by background dwarf stars would reduce it. The strong fluctuations of disc fraction as a function of mass seen in these samples will be caused primarily by the varying influence of these contaminating samples, rather than actual changes in the disc fraction of VLMOs in Upper Scorpius. Discrepancies on the brown dwarf disc fraction presented in the literature may to a large extent be caused by differences in sample selection.

Second, the clean samples with the \HKcut cut do show that disc fractions within the sub-stellar domain increase with decreasing mass. This is seen in all three samples in the left panel of \Figure\ref{figure:disc_as_function_of_mass}. This trend was already stated in previous work, most notably by \citet{Luhman2012b}. In our samples, the disc fraction for $M<$0.05\Msun is about 2-3 times larger than at $0.05<M<0.15$\Msun. This is solid evidence for disc lifetimes significantly exceeding 10$\,Myr$ for low-mass brown dwarfs, confirming previous claims based on smaller samples \citep{Riaz2009}.

\subsection{Mass function}
\label{section:properties:mass_function}

Using the same MCMC process as in \Section\ref{section:properties:discs} we worked out the total number of objects in each mass bin. The mass function was then calculated using the total number of objects in each mass bin divided by the size of the mass bin. Uncertainties in number are assumed to be $\sqrt{N}$, and the uncertainty on $dM$ is assumed to be the size of the mass bin. This means the uncertainty on $dN/dM$ is dominated by the uncertainty on the mass estimates. In \Figure\ref{figure:disc_as_function_of_mass} we plot the mass function with the median uncertainties shown in yellow and the mass functions from \citet{Kroupa2001}, $\alpha=1.3$ for $0.08<M<0.5$\Msun, $\alpha=0.3$ for $M<0.08$\Msun, scaled arbitrarily to match our number of objects. Note that $N$, and thus $dN/dM$, are 10,000 times higher than our samples due to the Monte-Carlo samples used. 

From \Figure\ref{figure:disc_as_function_of_mass} we can see that both the C-ZYJHK DR10 samples are a good match to the Kroupa IMF (between $\sim 0.01$ and $\sim0.1$\Msun, with best fits values of $\alpha=$0.45 and 0.38 for C-ZYJHK\noRedcut\HKcut and C-ZYJHK\Redcut\HKcut respectively). For the higher mass objects our brightness cuts impose mass cuts that are seen as the steep decrease in the number of objects with masses above 0.1\Msun. At low masses (especially in the case of the UKIDSS science verification data) we see a gradual decrease in numbers of objects due to fainter objects being missed by UKIDSS (more so in the science verification data due to its preliminary shallow nature). Overall for the small area not affected by the brightness cuts or the faintness limit our data is consistent with the Kroupa IMF. A value of $\alpha \sim 0.4$ is also in line with previous determinations of the IMF in this region, \eg, \citet{Lodieu2013b} find $\alpha =0.45 \pm 0.11$.

\begin{table*}
\begin{center}
\caption{The number of outliers (those with excess greater than 2 sigma) compared to the number of expected outliers from a Gaussian distribution (again beyond 2 sigma). \label{table:number_of_outliers}}
\begin{tabular}{lcccc}
\hline
\hline
Sample & Total objects in sample & Non-outliers & Outliers & Expected Outliers from Gaussian \\
\hline
L-ZYJHK 						& 415 & 347 & 68 & 57 \\
\hline
C-ZYJHK DR10\noRedcut\HKcut		& 49 & 43 & 6 & 7 \\
C-ZYJHK DR10\Redcut\HKcut		& 68 & 59 & 9 & 10 \\
C-ZYJHK SV\HKcut				& 14 & 10 & 4 & 2 \\
\hline
C-ZYJHK DR10\noRedcut			& 171 & 134 & 37 & 22 \\
C-ZYJHK DR10\Redcut				& 224 & 187 & 37 & 30 \\
C-ZYJHK SV						& 58 & 44 & 14 & 8 \\
\hline
\hline
\end{tabular}\end{center}
\end{table*}

The right panel in \Figure\ref{figure:disc_as_function_of_mass} which shows the samples without the \HKcut cut again illustrates the effects of contamination on the mass function. In particular, there is a consistent `bump' in the mass function just above 0.1\Msun, which is most likely introduced by background objects. Echoing our previous comments, we would like to caution using this selection method for candidate members to derive population statistics for $M>0.1$\Msun, unless comprehensive spectroscopic characterisation is carried out to confirm youth and membership.

\subsection{Proper motion outliers}
\label{section:properties:outliers}

In this subsection we explore the possibility that some of the VLMOs in Upper Scorpius have a dynamic history that deviates from the bulk of the population, for example, because they experienced an ejection in the early stages of their evolution, which might have stopped in-fall and constrained the mass. Ejections like that are part of a number of proposed formation scenarios for sub-stellar objects \citep[\eg][]{Whitwort2007}. 

To test for this possibility, we compare the proper motion for each target compared to its ten nearest neighbors (nearest in Right Ascension and Declination). An excess in proper motion was calculated for each object and is defined in \Equation\ref{equation:excess_pm}.

\begin{equation}
\label{equation:excess_pm}
\begin{split}
& \text{Excess }(\mu_{\alpha}) = \frac{(\mu_{\alpha})_{\text{Target}} - (\mu_{\alpha})_{\text{NN}}}
{\sqrt{\sigma^2_{(\mu_\alpha)_{\text{Target}}} + \sigma^2_{(\mu_{\alpha})_{\text{NN}}}}} \\
& \\
& \text{Excess }(\mu_{\delta}) = \frac{(\mu_{\delta})_{\text{Target}} - (\mu_{\delta})_{\text{NN}}}
{\sqrt{\sigma^2_{(\mu_\delta)_{\text{Target}}} + \sigma^2_{(\mu_{\delta})_{\text{NN}}}}} \\
\end{split}
\end{equation}

\noindent where $(\mu_{\alpha,\delta})_{\text{NN}}$ is the median of the ten nearest neighbor objects proper motion in the Right Ascension/Declination direction, and $\sigma_{(\mu_{\alpha,\delta})_{\text{NN}}}$ is the standard deviation of the ten nearest neighbor objects in the Right Ascension/Declination direction. Thus excess is in units of sigma, where a value of zero would equate to an object having a proper motion component exactly equivalent to that of it's neighbors.

This allowed us to flag any candidates with excesses beyond 2 sigma (defined by a circle in proper motion space). For the L-ZYJHK and L-HK only sample we found 68/415 and 31/175 outlier in our \UpperSco candidates. For the C-ZYJHK DR10\noRedcut\HKcut, C-ZYJHK DR10\Redcut\HKcut and C-ZYJHK SV\HKcut samples we found 6/49, 9/68 and 4/14 outliers in our \UpperSco candidates. For the C-ZYJHK DR10\noRedcut\noHKcut, C-ZYJHK DR10\Redcut\noHKcut and C-ZYJHK SV\noHKcut samples we found 37/171, 37/224, 14/58 outliers in our \UpperSco candidates. \Figure\ref{figure:pm_outliers} shows the excess distribution in proper motion for two samples, with the 2 sigma circle drawn and outliers highlighted in red. The numbers are reported in \Table\ref{table:number_of_outliers}.

Comparing the number of outliers with the outliers expected from a Gaussian distribution (last column in \Table\ref{table:number_of_outliers}), there is currently no strong evidence for the presence of a population of brown dwarfs with kinematical properties distinct from the bulk population in the same area. For samples without the \HKcut cut, the number of outliers is again expected to be affected by contamination, as discussed throughout this paper. We note that a proper motion of 2\,mas, currently the median uncertainty of our optimized proper motions, translates into a velocity in the plane of the sky of 1.3\,kms$^{-1}$, which is comparable with the typical velocity dispersion found in radial velocity surveys of young brown dwarfs \citep{Joergens2006}. Ejection velocities may in some cases be significantly beyond that level (\citealt{Stamatellos2009}, \citealt{Li2015}), \ie the lack of proper motion outliers already provides useful limits for formation scenarios. This topic is an area where we expect future data releases from Gaia to provide improved constraints.

\section{Concluding Remarks}
\label{section:summary}

The {\itshape Gaia} mission is a powerful new tool in understanding star forming regions due to its new and future astrometry. For the first time regions such as \UpperSco are no longer limited by the uncertainty on proper motion (and with future Gaia data releases unknown distances). As such we show that both age and mass estimates can be vastly improved, giving insight into the initial mass function and through published photometry from WISE, the disc fractions of these populations. However, we caution that future inferences on populations (such as for example, but in no way limited to, mass functions and disc fraction) will be very dependent on the selection criteria used. It may be that rigorous selection (\ie via full spectroscopic analysis) is required to really identify bona fide VLMOs and brown dwarfs from background objects and other contamination. This is especially valid for regions where reddening is important, but also applies, as shown in this paper, to regions free from extinction. Despite this we find that our mass functions (at least between $0.01<M<0.1\Msun$) are consistent with the Kroupa IMF, that the disc fraction among low-mass brown dwarfs ($M<0.05\Msun$) is substantially higher than in more massive objects. We note that proper motions from Gaia will give us an opportunity to detect  objects that were dynamically ejected early on. If done correctly, with future Gaia data and full spectroscopic follow-up the potential for nearby star forming regions to advance our knowledge of low-mass, very-low mass and planetary mass objects is extremely exciting.

\section*{Acknowledgments}

We would like to thank the anonymous referee whose careful reading of this paper and excellent comments were very welcome. We would like to thank Ogyen Verhagen, former undergraduate student at the University in St Andrews, whose final year project results motivated parts of this work. This work was supported in part by NSERC grants to RJ. This work has made use of data from the European Space Agency (ESA) mission {\it Gaia} (\url{https://www.cosmos.esa.int/gaia}), processed by the {\it Gaia} Data Processing and Analysis Consortium (DPAC, \url{https://www.cosmos.esa.int/web/gaia/dpac/consortium}). Funding for the DPAC has been provided by national institutions, in particular the institutions participating in the {\it Gaia} Multilateral Agreement. Specifically we use Gaia DR1 data (\citealt{Gaia2016a}, \citealt{Gaia2016b}) with more details available from \citet{Arenou2017}, \citet{Lindegren2016}, \citet{vanLeeuwen2017}, \citet{Carrasco2016} and \citet{Evans2017}. We make use of data products from WISE \citep{Wright2010}, which is a joint project of the UCLA, and the JPL$/$CIT, funded by NASA. The UKIDSS project is defined in \citet{Lawrence2007}. UKIDSS uses the UKIRT Wide Field Camera (WFCAM; \citealt{Casali2007}). The photometric system is described in \citet{Hewett2006}, and the calibration is described in \citet{Hodgkin2009}. The pipeline processing and science archive are described in \citet{Hambly2008}. We have used data from the DR10 data release and the science verification data, which are described in detail at \url{http://wsa.roe.ac.uk}. This work is based in part on services provided by the GAVO Data Center. This research has made use of the VizieR catalog access tool, CDS, Strasbourg, France \citep{Ochsenbein2000}. We acknowledge the use of data products from the HSOY catalog of \citet{Altmann2017}, the GPS1 catalog of \citet{Tian2017}, the UCAC5 catalog of \citet{Zacharias2017}, the Pan-STARRS1 catalog of \citet{Chambers2011} and the PPMXL database of \citet{Roeser2010}. This research has made use of NASA's Astrophysics Data System. 

\software{{\sc Astropy} \citep{Astropy2013}, 
          {\sc Ipython} \citep{Perez2007},
          {\sc Matplotlib} \citep{Barrett2005,Hunter2007}, 
          {\sc Numpy} \citep{jones2001,Oliphant2007},
          {\sc Scipy} \citep{jones2001,Oliphant2007},
          {\sc Stilts} \citep{Taylor2006},
          {\sc Topcat} \citep{Taylor2005},
          {\sc tqdm} \citep{costa_luis_2017}}



\appendix

\section{SQL Queries}
\label{section:sql_queries}

\subsection{The ZYJHK Sample}
\label{section:sql_queries:zyjhk}

ZYJHK sample query \citep[Private communications with N. Lodieu see][]{Lodieu2013a}. This query returned 2,653,897 sources from UKIDSS DR10 and 157,325 sources from the UKIDSS SV.

\begin{lstlisting}[language=SQL, literate={dec}{dec}3]
/* Start */
SELECT
    sourceID, ra, dec, zAperMag3, zAperMag3Err, yAperMag3, yAperMag3Err, jAperMag3, jAperMag3Err, hAperMag3, hAperMag3Err, k_1AperMag3, k_1AperMag3Err, muRa, muDec, sigMuRa, sigMuDec
FROM
    gcsSource
WHERE
       ra BETWEEN 232.0 AND 255.0
       AND dec BETWEEN -30.0 AND -15.0
       /* Bright saturation cut-offs */
       AND zaperMag3 > 11.3
       AND yaperMag3 > 11.5
       AND japerMag3 > 11.0
       AND haperMag3 > 11.30
       AND k_1aperMag3 > 9.90
       /* Limit merged passband selection to 1 arcsec */
       AND zXi BETWEEN -1.0 AND +1.0
       AND yXi BETWEEN -1.0 AND +1.0
       AND jXi BETWEEN -1.0 AND +1.0
       AND hXi BETWEEN -1.0 AND +1.0
       AND k_1Xi BETWEEN -1.0 AND +1.0
       AND zEta BETWEEN -1.0 AND +1.0
       AND yEta BETWEEN -1.0 AND +1.0
       AND jEta BETWEEN -1.0 AND +1.0
       AND hEta BETWEEN -1.0 AND +1.0
       AND k_1Eta BETWEEN -1.0 AND +1.0
       AND (jppErrBits < 131072)
       AND (hppErrBits < 131072)
       AND (k_1ppErrBits < 131072)
       /* Retain only point-like sources */
       AND (
        (
         ((zClass BETWEEN -2 AND -1) OR (zClassStat BETWEEN -3.0 AND +3.0))
        AND
         ((yClass BETWEEN -2 AND -1) OR (yClassStat BETWEEN -3.0 AND +3.0))
        AND
         ((jClass BETWEEN -2 AND -1) OR (jClassStat BETWEEN -3.0 AND +3.0))
        AND
         ((hClass BETWEEN -2 AND -1) OR (hClassStat BETWEEN -3.0 AND +3.0))
        AND
         ((k_1Class BETWEEN -2 AND -1) OR (k_1ClassStat BETWEEN -3.0 AND +3.0))
        )
        OR mergedClass BETWEEN -2 AND -1 OR mergedClassStat BETWEEN -3.0 AND +3.0
       )
       /* Retain only the best record when duplicated in an overlap region */
       AND (priOrSec = 0 OR priOrSec = frameSetID)
/* End */
\end{lstlisting}

\subsection{The HK-only Sample}
\label{section:sql_queries:hkonly}

HK-only sample query \citep[Private communications with N. Lodieu see][]{Lodieu2013a}. This query returned 7,473,530 sources from UKIDSS DR10.

\begin{lstlisting}[language=SQL, literate={dec}{dec}3]
/* Start */
SELECT
    sourceID, ra, dec, zAperMag3, zAperMag3Err, yAperMag3, yAperMag3Err, jAperMag3, jAperMag3Err, hAperMag3, hAperMag3Err, k_1AperMag3, k_1AperMag3Err, muRa, muDec, sigMuRa, sigMuDec
FROM
    gcsSource
WHERE
       ra BETWEEN 232.0 AND 255.0
       AND dec BETWEEN -30.0 AND -15.0
       /* Bright saturation cut-offs */
       AND (zaperMag3 < -0.9e9 OR zaperMag3 > 11.3)
       AND (yaperMag3 < -0.9e9 OR yaperMag3 > 11.5)
       AND (japerMag3 < -0.9e9 OR japerMag3 > 11.0)
       AND haperMag3 > 11.30
       AND k_1aperMag3 > 9.90
       /* Limit merged passband selection to 1 arcsec */
       AND (zXi BETWEEN -1.0 AND +1.0 OR zXi < -0.9e9)
       AND (yXi BETWEEN -1.0 AND +1.0 OR yXi < -0.9e9)
       AND (jXi BETWEEN -1.0 AND +1.0 OR jXi < -0.9e9)
       AND hXi BETWEEN -1.0 AND +1.0
       AND k_1Xi BETWEEN -1.0 AND +1.0
       AND (zEta BETWEEN -1.0 AND +1.0 OR zEta < -0.9e9)
       AND (yEta BETWEEN -1.0 AND +1.0 OR yEta < -0.9e9)
       AND (jEta BETWEEN -1.0 AND +1.0 OR jEta < -0.9e9)
       AND hEta BETWEEN -1.0 AND +1.0
       AND k_1Eta BETWEEN -1.0 AND +1.0
       AND (hppErrBits < 131072)
       AND (k_1ppErrBits < 131072)
       /* Retain only point-like sources */
       AND (
        (
         ((zClass BETWEEN -2 AND -1) OR (zClassStat BETWEEN -3.0 AND +3.0) OR (zClass = -9999))
        AND
         ((yClass BETWEEN -2 AND -1) OR (yClassStat BETWEEN -3.0 AND +3.0) OR (yClass = -9999))
        AND
         ((jClass BETWEEN -2 AND -1) OR (jClassStat BETWEEN -3.0 AND +3.0) OR (jClass = -9999))
        AND
         ((hClass BETWEEN -2 AND -1) OR (hClassStat BETWEEN -3.0 AND +3.0))
        AND
         ((k_1Class BETWEEN -2 AND -1) OR (k_1ClassStat BETWEEN -3.0 AND +3.0))
        )
        OR mergedClass BETWEEN -2 AND -1 OR mergedClassStat BETWEEN -3.0 AND +3.0
       )
       /* Retain only the best record when duplicated in an overlap region */
       AND (priOrSec = 0 OR priOrSec = frameSetID)
/* End */
\end{lstlisting}

\section{The samples by number}
\label{section:appendix_samples}

In tables \ref{table:samples_by_numbers_l} and \ref{table:samples_by_numbers_c} we show the source counts (\ie the number of objects used from the original tables), the number of objects with proper motions in the HSOY, GPS1, UCAC, GCS, and PPMXL catalogs. We show the number of objects with a `most precise' proper motion (in the `best pm' column, \ie the smallest uncertainty as chosen from the HSOY, GPS1, UCAC and GCS catalogs, see \Section\ref{section:pm_analysis}). We show the source counts for those object that have WISE photometry, have a disc as indicated by the Dawson cuts (`W1W2 disc', `W3 disc' and the combination of the two `disc' column, see \Section\ref{section:properties:discs}). The `mass est.' column describes whether we were able to find a isochronal model that fit the data (see \Section\ref{section:properties:isofit}). The `USco' column gives the number in each which conforms to our selection criteria for an \UpperSco member (see \Section\ref{section:pm_analysis}) and the numbers of objects in \UpperSco with discs, with a isochronal mass estimate and with both a disc and a isochronal mass estimate are shown in the last three columns. Tables \ref{table:csample_cols} and \ref{table:lsample_cols} shows the column descriptions for the samples (tables available online in machine readable format).

\begin{longrotatetable}
\begin{table}
\caption{The source counts for the L-sample. For the L-sample numbers are also presented by original literature source catalog and how many sources are present in UKIDSS GCS DR10 and the UKIDSS GCS Science verification data. Tables available online in machine readable format. \label{table:samples_by_numbers_l}}

\begin{tabular}{p{2.75cm}p{.75cm}p{.75cm}p{.75cm}p{.75cm}p{.75cm}p{1cm}p{.75cm}p{.75cm}p{.75cm}p{.75cm}p{.75cm}p{.75cm}p{.75cm}p{1.25cm}p{1.5cm}p{1.5cm}}

\multicolumn{17}{l}{}\\

\hline
\hline
Flag & Total & HSOY & GPS1 & UCAC & GCS & PPMXL & best pm & WISE & W1W2 Disc & W3 Disc & Disc & mass est. & USco & USco+ disc & USco+ Wise+ mass est. & USco disc+ mass est. \\
\hline
In D11 & 28 & 12 & 21 & 0 & 28 & 23 & 28 & 12 & 1 & 0 & 1 & 28 & 20 & 1 & 8 & 1 \\
In L11 & 91 & 63 & 77 & 36 & 91 & 83 & 91 & 62 & 6 & 14 & 17 & 91 & 88 & 16 & 59 & 16 \\
In D13 & 108 & 53 & 80 & 1 & 108 & 86 & 108 & 52 & 11 & 5 & 11 & 108 & 104 & 11 & 49 & 11 \\ 
In L13a ZYJHK\noRedcut & 201 & 124 & 154 & 36 & 201 & 159 & 201 & 124 & 13 & 12 & 17 & 184 & 195 & 17 & 107 & 15 \\ 
In L13a ZYJHK\Redcut & 120 & 73 & 84 & 23 & 120 & 96 & 120 & 72 & 6 & 4 & 8 & 116 & 108 & 7 & 59 & 7 \\
In L13a HK & 250 & 180 & 191 & 29 & 250 & 225 & 250 & 176 & 32 & 18 & 32 & 9 & 184 & 30 & 2 & 0 \\
In L13a ZYJHK SV & 70 & 38 & 55 & 5 & 70 & 61 & 70 & 38 & 6 & 4 & 8 & 66 & 64 & 7 & 33 & 7 \\ 
In L13b & 25 & 9 & 7 & 0 & 25 & 13 & 25 & 9 & 4 & 2 & 4 & 25 & 22 & 3 & 7 & 3 \\
In D14 & 30 & 14 & 24 & 0 & 30 & 25 & 30 & 14 & 4 & 3 & 4 & 30 & 23 & 4 & 9 & 4 \\ 
In GCS SV & 122 & 79 & 103 & 38 & 122 & 110 & 122 & 77 & 10 & 17 & 21 & 117 & 112 & 20 & 70 & 20 \\
In GCS DR10 & 716 & 469 & 549 & 126 & 716 & 609 & 716 & 462 & 66 & 55 & 83 & 453 & 610 & 79 & 244 & 47 \\ 
\hline
L-ZYJHK & 453 & 276 & 345 & 93 & 453 & 372 & 453 & 273 & 33 & 36 & 49 & 453 & 415 & 47 & 244 & 47 \\
L-HK only & 241 & 178 & 187 & 29 & 241 & 220 & 241 & 174 & 32 & 18 & 32 & 0 & 175 & 30 & 0 & 0 \\
\hline
Total & 716 & 469 & 549 & 126 & 716 & 609 & 716 & 462 & 66 & 55 & 83 & 453 & 610 & 79 & 244 & 47 \\
\hline
\hline
\end{tabular}
\end{table}

Notes: \\
\begin{itemize}
\item Source catalogs are as follows: D11 \citet{Dawson2011}; L11 \citet{Lodieu2011}; D13  \citet{Dawson2013}; L13a \citet{Lodieu2013a}; L13b \citet{Lodieu2013b}; D14 \citet{Dawson2014}; and GCS \citet{Lawrence2007} - the UKIDSS GCS catalog.
\item ZYJHK, R, and HK here is as in \citet{Lodieu2013a} not as in our definitions of ZYJHK, \Redcut\, or \HKcut.
\item SV denotes UKIDSS GCS science verification data, DR10 denotes data release 10 data.
\end{itemize}

\end{longrotatetable}

\begin{longrotatetable}
\begin{table}
\caption{The source counts for the C-sample. Tables available online in machine readable format. \label{table:samples_by_numbers_c}}

\begin{tabular}{p{2.75cm}p{.75cm}p{.75cm}p{.75cm}p{.75cm}p{.75cm}p{1cm}p{.75cm}p{.75cm}p{.75cm}p{.75cm}p{.75cm}p{.75cm}p{.75cm}p{1.25cm}p{1.5cm}p{1.5cm}}

\multicolumn{17}{l}{}\\
\multicolumn{17}{l}{The C-sample with the HK cut applied.}\\
\multicolumn{17}{l}{}\\

\hline
\hline
Flag & Total & HSOY & GPS1 & UCAC & GCS & PPMXL & best pm & WISE & W1W2 Disc & W3 Disc & Disc & mass est. & USco & USco+ disc & USco+ Wise+ mass est. & USco disc+ mass est. \\
\hline
C-ZYJHK DR10\noRedcut\HKcut & 66 & 21 & 24 & 0 & 66 & 36 & 66 & 63 & 14 & 3 & 15 & 66 & 49 & 14 & 46 & 14 \\
C-ZYJHK DR10\Redcut\HKcut & 77 & 24 & 35 & 0 & 77 & 42 & 77 & 73 & 12 & 1 & 12 & 77 & 68 & 11 & 64 & 11 \\
C-ZYJHK SV \HKcut & 17 & 12 & 14 & 0 & 17 & 15 & 17 & 17 & 5 & 0 & 5 & 17 & 14 & 4 & 14 & 4 \\
\hline
C-ZYJHK Total\HKcut & 160 & 57 & 73 & 0 & 160 & 93 & 160 & 153 & 31 & 4 & 32 & 160 & 131 & 29 & 124 & 29 \\
C-HK only & 1526 & 870 & 807 & 225 & 1526 & 1032 & 1519 & 1395 & 180 & 98 & 214 & 1526 & 346 & 88 & 331 & 88 \\
\hline
\hline

\multicolumn{17}{l}{}\\
\multicolumn{17}{l}{The C-sample without the HK cut applied.} \\
\multicolumn{17}{l}{}\\

\hline
\hline
Flag & Total & HSOY & GPS1 & UCAC & GCS & PPMXL & best pm & WISE & W1W2 Disc & W3 Disc & Disc & mass est. & USco & USco+ disc & USco+ Wise+ mass est. & USco disc+ mass est. \\
\hline
C-ZYJHK DR10\noRedcut\noHKcut & 1305 & 1212 & 1177 & 1063 & 1305 & 1262 & 1305 & 1281 & 21 & 25 & 42 & 1305 & 171 & 21 & 167 & 21 \\
C-ZYJHK DR10\Redcut\noHKcut & 811 & 688 & 728 & 546 & 811 & 752 & 811 & 797 & 24 & 19 & 33 & 811 & 224 & 29 & 217 & 29 \\
C-ZYJHK SV\noHKcut & 86 & 43 & 64 & 8 & 68 & 65 & 68 & 67 & 14 & 11 & 18 & 68 & 58 & 17 & 57 & 17 \\
\hline
C-ZYJHK Total\noHKcut & 2202 & 1943 & 1969 & 1617 & 2184 & 2079 & 2184 & 2145 & 59 & 55 & 93 & 2184 & 453 & 67 & 441 & 67 \\
\hline
\hline
\end{tabular}
\end{table}
\end{longrotatetable}

\startlongtable
\begin{deluxetable}{lllr}
\tablecaption{Column descriptions for C-sample tables for C-ZYJHK DR10\noRedcut\noHKcut, C-ZYJHK DR10\Redcut\noHKcut, C-ZYJHK SV\noHKcut, C-ZYJHK DR10\noRedcut\HKcut, C-ZYJHK DR10\Redcut\HKcut, and C-ZYJHK SV\HKcut and tables are available online in machine readable format. \label{table:csample_cols}}
\tablewidth{700pt}
\tabletypesize{\scriptsize}
\tablehead{
\colhead{Column Name} & \colhead{Description} & \colhead{Unit} & \colhead{UCD}
} 
\startdata
  UID & Unique identifier (1) & \nodata & meta.id;meta.main\\
  RAdeg & Right Ascension in decimal degrees (J2000) & deg & pos.eq.ra;meta.main\\
  DEdeg & Declination in decimal degrees (J2000) & deg & pos.eq.dec;meta.main\\
  Zmag & UKIDSS Z magnitude & mag & phot.mag;em.IR\\
  e\_Zmag & uncertainty in Zmag & mag & stat.error;em.IR\\
  Ymag & UKIDSS Y magnitude & mag & phot.mag;em.IR\\
  e\_Ymag & uncertainty in Ymag & mag & stat.error;em.IR\\
  Jmag & UKIDSS J magnitude & mag & phot.mag;em.IR.J\\
  e\_Jmag & uncertainty in Jmag & mag & stat.error;em.IR.J\\
  Hmag & UKIDSS H magnitude & mag & phot.mag;em.IR.H\\
  e\_Hmag & uncertainty in Hmag & mag & stat.error;em.IR.H\\
  Kmag & K magnitude & mag & phot.mag;em.IR.K\\
  e\_Kmag & uncertainty in Kmag & mag & stat.error;em.IR.K\\
  pm & Most precise proper motion & mas/yr & pos.pm\\
  e\_pm & Uncertainty in pm & mas/yr & stat.error;pos.pm\\
  r\_pm & Reference for pm (2) & \nodata & meta.bib\\
  pmRA & Most precise proper motion in RA & mas/yr & pos.pm;pos.eq.ra\\
  e\_pmRA & Uncertainty in pmRA & mas/yr & stat.error;pos.pm;pos.eq.ra\\
  pmDE & Most precise proper motion in DE & mas/yr & pos.pm;pos.eq.dec\\
  e\_pmDE & Uncertainty in pmDE & mas/yr & stat.error;pos.pm;pos.eq.dec\\
  WISE & Has WISE photometry (3) & \nodata & code.meta\\
  Disk & Flagged as having a disc (3) (4) & \nodata & code.meta\\
  MassFit & Best $\chi^{2}$ fit for mass & solMass & phys.mass\\
  bMassFit & Lower uncertainty bound in MassFit & solMass & stat.error;phys.mass\\
  BMassFit & Upper uncertainty bound in MassFit & solMass & stat.error;phys.mass\\
  TeffFit & Best $\chi^{2}$ fit for effective temperature & K & phys.temperature.effective\\
  bTeffFit & Lower uncertainty bound in TeffFit & K & stat.error;phys.temperature.effective\\
  BTeffFit & Upper uncertainty bound in TeffFit & K & stat.error;phys.temperature.effective\\
  LumFit & Best $\chi^{2}$ fit for luminosity & solLum & phys.luminosity\\
  bLumFit & Lower uncertainty bound in LumFit & solLum & stat.error;phys.luminosity\\
  BLumFit & Upper uncertainty bound in LumFit & solLum & stat.error;phys.luminosity\\
  log(g)Fit & Best $\chi^{2}$ fit for surface gravity & [cm/s2] & phys.gravity\\
  blog(g)Fit & Lower uncertainty bound in log(g)Fit & [cm/s2] & stat.error;phys.gravity\\
  Blog(g)Fit & Upper uncertainty bound in log(g)Fit & [cm/s2] & stat.error;phys.gravity\\
  RadFit & Best $\chi^{2}$ fit for radius & solRad & phys.size.radius\\
  bRadFit & Lower uncertainty bound in RadFit & solRad & stat.error;phys.size.radius\\
  BRadFit & Upper uncertainty bound in RadFit & solRad & stat.error;phys.size.radius\\
  chilow & Low fit $\chi^{2}$ & \nodata & stat.fit.chi2\\
  chimid & Mid fit $\chi^{2}$ & \nodata & stat.fit.chi2\\
  chihi & High fit $\chi^{2}$ & \nodata & stat.fit.chi2\\
  WiseAndMass & Has WISE photometry and a mass estimate & \nodata & meta.code\\
  sigExpmRA & Sigma excess in pmRA & \nodata & stat.value;pos.pm;pos.eq.ra\\
  sigExpmDE & Sigma excess in pmDE & \nodata & stat.value;pos.pm;pos.eq.dec\\
  fnpmout & Not an outlier in pm; 2$\sigma$ (3) & \nodata & meta.code.member\\
  fpmout & Is an outlier in pm; 2$\sigma$ (3) & \nodata & meta.code.member\\
\enddata
\tablenotetext{1}{Of this merged detection as assigned by merge algorithm. ID is unique over entire WSA via program ID prefix.}
\tablenotetext{2}{hsoy = Hot Stuff for One Year (HSOY) catalog (\citealt{Altmann2017}); gps1 = Gaia-PS1-SDSS (GPS1) catalog (\citealt{Tian2017}); gcs = The UKIRT Infrared Deep Sky Survey (UKIDSS) catalog (\citealt{Lawrence2007}); and ucac = UCAC5: New Proper Motions Using Gaia DR1 catalog (\citealt{Zacharias2017}).}
\tablenotetext{3}{1 = True, 2 = False}
\tablenotetext{4}{\citet{Dawson2013} WISE W1-W2 or WISE W3 disc.}
\end{deluxetable}

\vspace{1cm}

\startlongtable
\begin{deluxetable}{lllr}
\tablecaption{Column descriptions for the L-sample tables (L-ZJYHK and L-HKonly). Tables available online in machine readable format. \label{table:lsample_cols}}
\scriptsize
\tablewidth{700pt}
\tabletypesize{\scriptsize}
\tablehead{
\colhead{Column Name} & \colhead{Description} & \colhead{Unit} & \colhead{UCD}
} 
\startdata
  UID & Unique ID from catalog creation & \nodata & meta.id;meta.main\\
  RAdeg & Right Ascension in decimal degrees (J2000) & deg & pos.eq.ra;meta.main\\
  DEdeg & Declination in decimal degrees (J2000) & deg & pos.eq.dec;meta.main\\
  Coord & Source of main coordinates & \nodata & meta.bib\\
  Cat & Catalog contained in (1) & \nodata & meta.bib\\
  Zmag & Most precise Z magnitude & mag & phot.mag;em.opt.Z\\
  r\_Zmag & Reference for Zmag (1) & \nodata & meta.bib\\
  e\_Zmag & uncertainty in Zmag & mag & stat.error;em.opt.Z\\
  Ymag & Most precise Y magnitude & mag & phot.mag;em.opt.Y\\
  r\_Ymag & Reference for Ymag (1) & \nodata & meta.bib\\
  e\_Ymag & uncertainty in Ymag & mag & stat.error;em.opt.Y\\
  Jmag & Most precise J magnitude & mag & phot.mag;em.IR.J\\
  r\_Jmag & Reference for Jmag (1) & \nodata & meta.bib\\
  e\_Jmag & uncertainty in Jmag & mag & stat.error;phot.mag;em.IR.J\\
  Hmag & Most precise H magnitude & mag & phot.mag;em.IR.H\\
  r\_Hmag & Reference for Hmag (1) & \nodata & meta.bib\\
  e\_Hmag & uncertainty in Hmag & mag & stat.error;phot.mag;em.IR.H\\
  Kmag & Most precise K magnitude & mag & phot.mag;em.IR.K\\
  r\_Kmag & Reference for Kmag (1) & \nodata & meta.bib\\
  e\_Kmag & uncertainty in Kmag & mag & stat.error;phot.mag;em.IR.K\\
  SpTCat & spt from source catalog & \nodata & src.spType\\
  r\_SpTCat & Reference for SpTcat (1) & \nodata & meta.bib\\
  MassCat & Mass from source catalog & solMass & phys.mass\\
  r\_MassCat & Reference for MassCat (1) & \nodata & meta.bib\\
  fZYJHK & Flag whether source ahs valid ZYJHK photometry (2) & \nodata & meta.code\\
  fHKonly & Flag that source only has HK photometry (2) & \nodata & meta.code\\
  pm & Most precise proper motion & mas/yr & pos.pm\\
  e\_pm & Uncertainty in pm & mas/yr & stat.error;pos.pm\\
  r\_pm & Reference for pm (3) & \nodata & meta.bib\\
  pmRA & Most precise proper motion in RA & mas/yr & pos.pm;pos.eq.ra\\
  e\_pmRA & Uncertainty in pmRA & mas/yr & stat.error;pos.pm;pos.eq.ra\\
  pmDE & Most precise proper motion in DE & mas/yr & pos.pm;pos.eq.dec\\
  e\_pmDE & Uncertainty in pmDE & mas/yr & stat.error;pos.pm;pos.eq.dec\\
  MassFit & Best $\chi^{2}$ fit for mass & solMass & phys.mass\\
  bMassFit & Lower uncertainty bound in MassFit & solMass & stat.error;phys.mass\\
  BMassFit & Upper uncertainty bound in MassFit & solMass & stat.error;phys.mass\\
  TeffFit & Best $\chi^{2}$  fit for effective temperature & K & phys.temperature.effective\\
  bTeffFit & Lower uncertainty bound in TeffFit & K & stat.error;phys.temperature.effective\\
  BTeffFit & Upper uncertainty bound in TeffFit & K & stat.error;phys.temperature.effective\\
  LumFit & Best $\chi^{2}$  fit for luminosity & solLum & phys.luminosity\\
  bLumFit & Lower uncertainty bound in LumFit & solLum & stat.error;phys.luminosity\\
  BLumFit & Upper uncertainty bound in LumFit & solLum & stat.error;phys.luminosity\\
  log(g)Fit & Best $\chi^{2}$  fit for surface gravity & [cm/s2] & phys.gravity\\
  blog(g)Fit & Lower uncertainty bound in log(g)Fit & [cm/s2] & stat.error;phys.gravity\\
  Blog(g)Fit & Upper uncertainty bound in log(g)Fit & [cm/s2] & stat.error;phys.gravity\\
  RadFit & Best $\chi^{2}$  fit for radius & solRad & phys.size.radius\\
  eRadFitL & Lower uncertainty bound in RadFit & solRad & stat.error;phys.size.radius\\
  eRadFitU & Upper uncertainty bound in RadFit & solRad & stat.error;phys.size.radius\\
  chilow & Low fit $\chi^{2}$  & \nodata & stat.fit.chi2\\
  chimid & Mid fit $\chi^{2}$  & \nodata & stat.fit.chi2\\
  chihi & High fit $\chi^{2}$  & \nodata & stat.fit.chi2\\
  Disk & Flagged as having a disc (2) (4) & \nodata & meta.code\\
  WiseAndMass & Has WISE photometry and a mass estimate (2) & \nodata & meta.code\\
  sigExpmRA & Sigma excess in pmRA & \nodata & stat.value;pos.pm;pos.eq.ra\\
  sigExpmDE & Sigma excess in pmDE & \nodata & stat.value;pos.pm;pos.eq.dec\\
  fnotPMout & Not an outlier in pm; 2$\sigma$ (2) & \nodata & meta.code.member\\
  fPMout & Is an outlier in pm; 2$\sigma$ (2) & \nodata & meta.code.member\\
\enddata\tablenotetext{1}{GCSSV = from UKIDSS Science verification (\citealt{Lawrence2007}); GCSDR10 = from UKIDSS DR10 (\citealt{Lawrence2007}); D11 = from \citet{Dawson2011}]; L11 = From \citet{Lodieu2011}; D13 = from \citet{Dawson2013}; L13a = from \citet{Lodieu2013a}: L13a\_HK from HK sample, L13a\_ZYJHK\_SV from the ZYJHK UKIDSS Science verification sample,L13a\_ZYJHK\_red from the ZYJHK UKIDSS DR10 affected by reddening sample, L13a\_ZYJHK\_nored from the ZYJHK UKIDSS DR10 devoid of reddening sample; L13b = from \citet{Lodieu2013b}; D14 = from \citet{Dawson2014}.}
\tablenotetext{2}{1 = True, 2 = False}
\tablenotetext{3}{hsoy = Hot Stuff for One Year (HSOY) catalog (\citealt{Altmann2017}); gps1 = Gaia-PS1-SDSS (GPS1) catalog (\citealt{Tian2017}); gcs = The UKIRT Infrared Deep Sky Survey (UKIDSS) catalog (\citealt{Lawrence2007}); and ucac = UCAC5: New Proper Motions Using Gaia DR1 catalog (\citealt{Zacharias2017}).}
\tablenotetext{4}{From isochrones (i.e have ZYJHKW1W2 photometry).}
\end{deluxetable}






\end{document}